\documentclass[12pt,preprint,letterpaper,nofootinbib,superscriptaddress]{revtex4}
\pdfoutput=1
\usepackage[T1]{fontenc}
\usepackage[utf8]{inputenc}
\usepackage{epsf, array, xcolor}
\usepackage{graphicx}
\usepackage{subfigure}
\usepackage{bbold}
\usepackage{amsmath,amssymb,mathtools}
\usepackage{epigraph}
\usepackage[colorlinks=true,linktocpage=true,citecolor=blue,linkcolor=blue,urlcolor=blue]{hyperref}
\usepackage{textcomp}
\usepackage{color}
\usepackage{slashed}
\usepackage{graphics,psfrag}
\usepackage{graphicx,psfrag}
\usepackage{cancel}
\usepackage{comment}
\usepackage{url}
\usepackage{dcolumn}
\usepackage{bm}

\begin{document}
\pagestyle{plain}

\title{SMEFT Matching to $Z^\prime$ Models at Dimension-8}

\author{Sally Dawson}
\affiliation{Department of Physics, Brookhaven National Laboratory, Upton, N.Y., 11973,  U.S.A.}
\author{Matthew Forslund}
\affiliation{Department of Physics, Brookhaven National Laboratory, Upton, N.Y., 11973,  U.S.A.}
\affiliation{C. N. Yang Institute for Theoretical Physics, Stony Brook University, Stony Brook, N.Y., 11794, U.S.A.}
\author{Marvin Schnubel}
\affiliation{Department of Physics, Brookhaven National Laboratory, Upton, N.Y., 11973,  U.S.A.}

\date{\today}

\begin{abstract}
Heavy neutral gauge bosons arise in many motivated models of Beyond the Standard Model Physics. 
Experimental searches require that such gauge bosons are above the TeV scale in most models which means that the tools of effective field theories, in particular the Standard Model Effective Field Theory (SMEFT), are useful.  
We match the SMEFT to models with heavy $Z^\prime$ bosons, including effects of dimension-8 operators, and consider the restrictions on model parameters from electroweak precision measurements and from Drell Yan  invariant mass distributions and forward-backward asymmetry, $A_\text{FB}$, measurements at the LHC.  
The results demonstrate the model dependence of the resulting limits on SMEFT coefficients and the relatively small impact of including dimension-8 matching.  
In all cases, the limits from invariant mass distributions are stronger than from $A_\text{FB}$ measurements in the $Z^\prime$ models we consider. 
\end{abstract} 

\maketitle

\section{Introduction} 
The search for new physics at the LHC is one of the major efforts of the high luminosity run.  New physics can manifest itself through the discovery of new particles or in deviations from Standard Model (SM) predictions. Precision measurements are often sensitive to very high mass scales above those that can be probed by direct particle production, motivating an effective field theory program of probing for small differences from SM predictions.  In the scenario when no new particles are discovered, the SM effective field theory (SMEFT)~\cite{Buchmuller:1985jz} is a useful tool where new physics is described by an effective Lagrangian $\mathcal{L}$ that is an expansion around the SM,
\begin{equation}
 \mathcal{L}=\mathcal{L}_\text{SM}+\sum_{\alpha,d>4}{C_\alpha^{(d)}{\mathcal{O}}^{(d)}_\alpha\over \Lambda^{d-4} } .
 \label{eq:lsmeft}
\end{equation}
Any new physics is entirely contained in the coefficient functions, $C_\alpha^{(d)}$, and  the operators ${\mathcal{O}}_\alpha^{(d)}$ contain only SM particles and respect the SM $SU(3)\times SU(2)_L\times U(1)_Y$ gauge symmetry.  The expansion is then in inverse powers of a heavy scale $\Lambda$.  The odd dimension operators violate lepton number~\cite{Weinberg:1979sa} and will not be considered here.  A non-zero measurement of a coefficient, $C_\alpha^{(d)}$,
would be direct evidence for new physics and the goal of the LHC SMEFT program is to make precise measurements of these coefficients.

Experimental observables are often computed using the SMEFT Lagrangian truncated at dimension-6.  At this order, the only sensitivity is to 
the combination, ${C_i^{(6)} /\Lambda^2}$.   A complete basis for dimension-6 and dimension-8 operators exists~\cite{Ma:2019gtx,Aoude:2019tzn,Lehman:2014jma,Liao:2016hru,Murphy:2020rsh,Li:2020gnx,AccettulliHuber:2021uoa,Durieux:2019eor,Harlander:2023psl}, but there are too many operators for a general study of most processes to be feasible~\cite{Henning:2015alf,Henning:2017fpj,Fonseca:2019yya,Criado:2019ugp}, and typically only a small subset of the dimension-8 operators are studied~\cite{Hays:2018zze,Degrande:2023iob,Boughezal:2021tih,Boughezal:2022nof,Corbett:2023qtg}.  One approach that has proven useful is to match the SMEFT to a specific UV complete model with non-SM particles at the high scale $\Lambda$ which generates a relatively small number of operators whose effects can then be studied in low energy and weak scale processes. Different UV models have different patterns of coefficients and the hope is that by exploring these patterns, information about high scale physics can be obtained~\cite{Brivio:2017vri, Isidori:2023pyp,Dawson:2020oco,Gorbahn:2015gxa}.

The matching of dimension-6 operators to scenarios with a single new heavy particle is a solved problem, both at tree level and at one-loop~\cite{Fuentes-Martin:2020udw,Carmona:2021xtq,DasBakshi:2018vni,Criado:2017khh}.  The matching to more complicated models,  including dimension-8 contributions at tree level, has been performed for only  a small number of cases:  the two-Higgs doublet model~\cite{Dawson:2022cmu,Banerjee:2022thk}, singlet~\cite{Ellis:2023zim,Banerjee:2022thk,Banerjee:2023qbg}, top vector-like quark~\cite{Dawson:2021xei}, and scalar triplet models~\cite{Corbett:2021eux}.  Here we extend this program to include heavy neutral gauge bosons.

Heavy vector bosons are particularly interesting because they appear in many theoretically motivated extensions of the SM~\cite{Langacker:1984dc,Langacker:2008yv,Barger:2003zh,Babu:1987kp,Witten:1985bz,Hewett:1988xc,Robinett:1982tq,Antoniadis:1990ew,Appelquist:2000nn,Barbieri:2004qk,Gogoladze:2006br,Perelstein:2005ka,Chivukula:2003wj,Hill:2002ap,Agashe:2007ki,Carena:2003fx,Accomando:2010fz,Cvetic:1995zs,Rizzo:2006nw,Gulov:2018zij,Salvioni:2009mt,Chanowitz:2011ew,Accomando:2013sfa,Pappadopulo:2014qza,Baker:2022zxv}.  We match $Z'$ models with no SM hypercharge and  arbitrary $SU(3)\times SU(2)_L\times U(1)_Y$ invariant couplings to SM particles to the dimension-6 and dimension-8 SMEFT Lagrangians and study the resulting patterns of coefficients that arise. We also include an arbitrary kinetic mixing term in these models which turns out to have important phenomenological consequences.  Direct searches for heavy $Z^\prime$ models at the LHC give model dependent limits of ${\cal{O}}(2-5)$ TeV~\cite{ATLAS:2019erb,CMS:2021ctt}, suggesting that the SMEFT framework is applicable. Restrictions on heavy $Z'$ models matched to the SMEFT  come primarily from electroweak precision observables, along with Drell-Yan invariant mass distributions and forward backward asymmetries at the LHC and we study the interplay of these constraints and examine the sensitivity  to the specifics of the  $Z'$ models. 

The paper contains a description of  the $Z'$ models we consider in Section \ref{sec:zp} and presents results for the SMEFT matching to dimension-8 in Section \ref{sec:eft}.  The phenomenology of Drell-Yan production and the forward backward asymmetry at the LHC are compared with $Z$ pole observables in Section \ref{sec:res} and the numerical relevance of dimension-8 contributions to the computation of these observables is considered, along with the model dependence of our results. Section \ref{sec:con} contains our conclusions.

\section{Models with a heavy $Z^\prime$}
\label{sec:zp}

While there are many models extending the SM with new $U(1)^\prime$ symmetries, we  focus on a select few in this work and assume that the $Z^\prime$ is a singlet under all SM gauge groups. The models are distinguished by the charges of the SM particles under the new symmetry, and if the $U(1)^\prime$ is spontaneously broken, also by the breaking scale. An analysis of how to differentiate between  models in the case of a direct discovery at a future collider has recently been presented in Ref.~\cite{Korshynska:2024suh}. Note that in principle a mass term for the associated $Z^\prime$ boson of a $U(1)^\prime$ model  can also be generated via the Stückelberg mechanism~\cite{Stueckelberg:1938zz,Stueckelberg:1938hvi,Erler:2023eys,Ruegg:2003ps}. The $Z^\prime$ in this case
behaves like  a heavy dark photon and the distinction between the two is arbitrary and the $U(1)^\prime$ remains unbroken. Below we give a brief summary of  the models studied in this paper in order to set the notation.

\paragraph{Secluded $U(1)^\prime$:}
If the new symmetry belongs to a dark sector that is completely decoupled from the SM, the $Z^\prime$ interacts with SM particles only through kinetic mixing. This model is fully characterized by the strength of the kinetic mixing $\epsilon$ and the mass $M_{Z^\prime}$.

\paragraph{Hypercharge mirror:}
In hypercharge mirror models it is assumed that the additional $U(1)^\prime$ is a simple copy of the SM $U(1)_Y$ hypercharge gauge symmetry. The $U(1)^\prime$ charges of the SM fermions are therefore their respective hypercharges, and they couple to the $Z^\prime$ with the coupling constant $g_D$, which is often assumed to be comparable in size to the electroweak coupling constants, though smaller or larger values are not forbidden in general. The values considered in this work always respect the perturbativity limit, $g_D < 4\pi$. Note that we do not include an analysis of the sequential Standard Model (SSM) in our work. The SSM assumes that a copy of the whole SM gauge structure exists at a higher scale. Since the $Z^\prime$ is naturally accompanied by $W^\prime$ bosons of similar mass, their contributions would  need to be taken into account when matching the SSM onto SMEFT and examining phenomenological restrictions~\cite{Schmaltz:2010xr,Grojean:2011vu,Torre:2011wy,Barger:1980ix,Cvetic:1995zs}.

\paragraph{Gauging SM accidental symmetries:}
The SM features four accidental $U(1)$ symmetries, the individual lepton numbers ($L_e$, $L_\mu$ and $L_\tau$) and baryon number ($B$), respectively. However, the requirement of Adler--Bell--Jackiw anomaly cancellation dictates that only the differences of two lepton numbers $L_i-L_j$ or the difference of total baryon and lepton number $B-L$, with $L=L_e+L_\mu+L_\tau$, can be gauged in a consistent manner, assuming right-handed neutrinos are also added to the matter content. Under a $U(1)_{L_i-L_j}$ symmetry a lepton of generation $i$ has charge $+1$ and a lepton of generation $j$ has charge $-1$. In the $U(1)_{B-L}$ case all quarks have $U(1)_{B-L}$ charge $+1/3$ and the leptons have charge $-1$. All other particles are uncharged under the $U(1)$ symmetry in both cases. Note that for Drell-Yan processes a heavy $Z^\prime_{B-L}$ will, once integrated out of the theory, give rise to mixed quark-lepton four-fermion operators, while a $Z^\prime_{L_i-L_j}$ will only alter the couplings of the SM $Z$ boson to leptons.

\paragraph{Models based on $E_6$ symmetries:}
In the literature, $E_6$ models have played an important role as possible GUT symmetry candidates~\cite{Hewett:1988xc,Langacker:1984dc,Robinett:1982tq,Langacker:2008yv}. Their general breaking pattern is given by
\begin{equation}
	\label{eq:E6break}
	E_6\to SO(10)\times U(1)_\psi\to SO(5)\times U(1)_\psi\times U(1)_\chi\,.
\end{equation}
In $E_6$ models, the left-handed SM fermion families are promoted to a fundamental $\bm{ 27}$-plet, eventually decomposing into $\bm{27}\to(\bm{10}+\bm{5}^\ast+\bm{1})+(\bm{5}+\bm{5}^\ast)+\bm{1}$. Each $\bm{27}$-plet furthermore contains a conjugate of a right-handed neutrino $\nu^c$ and a new scalar $S$, both of which are singlets under the SM gauge groups. We will consider them as examples of anomaly-free models incorporating additional $U(1)$ symmetries, while being agnostic about the underlying grand unifying theory. Furthermore, we assume that there is only one $Z^\prime$ boson present corresponding to the linear combination of $E_6$ charges
\begin{equation}
	\label{eq:E6lin}
	Q_{E_6}=\cos\theta_{E_6}Q_\chi+\sin\theta_{E_6}Q_\psi\,,
\end{equation}
and $0\leq\theta_{E_6}\leq\pi$ is the mixing angle between the two $U(1)$ symmetries. Note that all $E_6$ models feature an implementation of two-Higgs doublet models (2HDM). Since the exact realization of the 2HDM affects the Higgs charges and couplings and hence the matching of the heavy $Z^\prime$ onto SMEFT operators, we will assume a type-I implementation of the 2HDM \cite{Branco:2011iw}. 

In previous works several benchmark models of $E_6$ have been studied. The $\psi$ and $\chi$ models assume that only one of the $U(1)$ symmetries is actually realized, and correspond to the mixing angles $\theta_\psi=\pi/2$ and $\theta_\chi=0$, respectively. The $\eta$ model is an example of $E_6$ directly breaking to $SO(5)$ via the Wilson line mechanism, and occurs in Calabi-Yau compactifications of the heterotic string~\cite{Witten:1985bz}. It corresponds to the mixing angle $\theta_\eta=\pi-\arctan\sqrt{5/3}$. The orthogonal case, where $E_6$ breaks to $SO(10)$ which is then directly broken to the SM gauge group, is called the inert model, and the mixing angle is $\theta_I=\arctan\sqrt{3/5}$~\cite{Robinett:1982tq}. In the neutral model with mixing angle $\theta_N=\arctan\sqrt{15}$, the right-handed neutrino $\nu^c$ has zero charge, hence allowing a large Majorana mass and the see-saw mechanism to take place~\cite{Barger:2003zh,Kang:2004ix,King:2005jy,Ma:1995xk}. Essentially the roles of the new scalar $S$ and $\nu^c$ are interchanged with regard to the $\chi$ model. Additionally, it can be regarded as an implementation of alternative left-right models~\cite{Babu:1987kp,Ma:1986we}. Lastly, the secluded sector model with mixing angle $\theta_S=\arctan(\sqrt{15}/9)$ was inspired by supersymmetric extensions of the SM~\cite{Erler:2002pr,Langacker:2008yv}.

In principle, other exotic models featuring heavy $Z^\prime$ realizations are feasible, like models where the $Z^\prime$ is a Kaluza-Klein excitation of a SM field~\cite{Antoniadis:1990ew,Appelquist:2000nn,Appelquist:2002wb,Barbieri:2004qk,Casalbuoni:1999ns,Cheng:2002ab,Cheung:2001mq,Delgado:1999sv,Gogoladze:2006br,Masip:1999mk,Davoudiasl:2007cy}, but they will be omitted in this work for simplicity. In Table~\ref{tab:charges}, we collect the $U(1)^\prime$ charges, $Q^f$,  of SM particles in the different models considered here, where $q^k_L$, $l^k_L$ are the left-handed fermion doublets, $H$ is the SM Higgs doublet, and $u^k_R,d^k_R$ and $l_R^k$ are the right-handed quarks and charged lepton, with $k$ a generation index\footnote{Depending on the model, $f^k_{R}$ may also include right handed neutrinos $\nu_R^k$. This is not relevant for the phenomenology we consider and so we ignore it.} . The $U(1)^\prime $ charges only depend on the generation for the leptons and we will often omit this index in the quark sector. 

Note that since a kinetic mixing of the $Z^\prime$ with the SM hypercharge gauge boson of strength $\epsilon$ is always possible, we will always take it into account in our results. 
In the dark matter literature, $\epsilon$ is often thought to be small, motivated in part by stringent experimental bounds (see for example~\cite{Fabbrichesi:2020wbt}). 
However, since we consider heavy $Z^\prime$ models, this is not necessarily the case for us. 
If we assume that $\epsilon=0$ at a high scale, which must be the case if the $U(1)^\prime$ or $U(1)_Y$ is embedded in a non-Abelian group such as in the $E_6$ models, $\epsilon$ will necessarily be generated by renormalization group evolution since the SM fermions are charged under both.
This can easily generate $\epsilon \sim \mathcal{O}(10^{-2}-1)$, depending on the values of the $U(1)^\prime$ gauge coupling, the charged matter content, and the relevant UV scale.
If the two $U(1)$ gauge groups are not embedded in any non-Abelian gauge group, then there is no a priori reason to take $\epsilon$ to be small even at the high scale.
Since we only have access to low energy information, we will take a bottom-up perspective, allowing for $\epsilon\sim 1$ in our results.

\begin{table}[t]
	\centering
	\begin{tabular}{||c|c|c|c|c|c|c|c|c|c||}
		\hline\hline
		SM particle & $Y_D$ & $B-L$ & $L_i-L_j$ & $E_6$, $\psi$ & $E_6$, $\chi$ & $E_6$, $\eta$ & $E_6$, inert & $E_6$, neutral & $E_6$, secluded \\
			\hline \hline
		$H$ & $\frac12$ & $0$ & $0$ & $\frac{-1}{\sqrt{6}}$ & $\frac{-1}{\sqrt{10}}$ & $\frac{1}{2\sqrt{15}}$ & $\frac12$ & $\frac{-3}{2\sqrt{10}}$ &  $\frac{-7}{4\sqrt{15}}$\\
		$q^k_L$ & $\frac16$ & $\frac13$ & $0$ & $\frac{1}{2\sqrt{6}}$ & $\frac{-1}{2\sqrt{10}}$ & $\frac{-1}{\sqrt{15}}$ & $0$ & $\frac{1}{2\sqrt{10}}$ & $\frac{-1}{4\sqrt{15}}$ \\
		$u^k_R$ & $\frac23$ & $\frac13$ & $0$ & $\frac{1}{2\sqrt{6}}$ & $\frac{-1}{2\sqrt{10}}$ & $\frac{-1}{\sqrt{15}}$ & $0$ & $\frac{1}{2\sqrt{10}}$ & $\frac{-1}{4\sqrt{15}}$\\
		$d^k_R$ & $-\frac{1}{3}$ & $\frac13$ & $0$ & $\frac{1}{2\sqrt{6}}$ & $\frac{3}{2\sqrt{10}}$ & $\frac{1}{2\sqrt{15}}$ & $-\frac{1}{2}$ & $\frac{1}{\sqrt{10}}$ & $\frac{2}{\sqrt{15}}$\\
		$l_L^k$ & $-\frac12$ & $-1$ & $\delta_{ik}-\delta_{jk}$ & $\frac{1}{2\sqrt{6}}$ & $\frac{3}{2\sqrt{10}}$ & $\frac{1}{2\sqrt{15}}$ & $-\frac{1}{2}$ & $\frac{1}{\sqrt{10}}$ &  $\frac{2}{\sqrt{15}}$\\
		$\ell^k_R$ & $-1$ & $-1$ & $\delta_{ik}-\delta_{jk}$ & $\frac{1}{2\sqrt{6}}$ & $\frac{-1}{2\sqrt{10}}$ & $\frac{-1}{\sqrt{15}}$ & $0$ & $\frac{1}{2\sqrt{10}}$ & $\frac{-1}{4\sqrt{15}}$\\
		\hline\hline
	\end{tabular}
	\caption{Charges, $Q^f$, of SM particles under the new gauge symmetry associated with the $Z^\prime$ corresponding to the models described in the text. $i, j, k$ are generation indices with $f_L^k=(q^k_L, l^k_L)$,  $(f_R^k=u^k_R, d^k_R, e^k_R)$. Here, $Y_D$ is the mirror hypercharge, hence the charges under this symmetry are exactly the SM hypercharges. In this table we omit the case of pure kinetic mixing, where all particles are uncharged under the new symmetry.
		\label{tab:charges}}
\end{table}

\section{SMEFT and Matching}
\label{sec:eft}
In this section, we review the basics of SMEFT that are relevant for the matching to $Z^\prime$ models.  We begin with the  SMEFT Lagrangian of Eq.~\eqref{eq:lsmeft} truncated at dimension-8.   The goal is  to compute the coefficient functions $C_i^{(6)}$ and $C_i^{(8)}$ that arise from integrating a heavy $Z^\prime$ out of the theory~\cite{Henning:2014wua}.
We add a real spin-1 boson $Z'$ that is a singlet under all SM gauge groups and consider the most general renormalizable Lagrangian,
\begin{equation}
    \mathcal{L}_{Z'} = - \frac{1}{4} Z'_{\mu\nu}Z'^{\mu\nu} + \frac{1}{2} M_{Z'}^2 Z'_\mu Z'^\mu + \frac{\epsilon}{2} B_{\mu\nu} Z'^{\mu\nu} + \left(g_{H,2} \right)^2 Z'_\mu Z'^\mu |H^\dagger H| - Z'_\mu \mathcal{J}^\mu ,
     \label{eq:lg}
\end{equation}
where $B^\mu$ is the $U(1)_Y$ hypercharge gauge field and $B_{\mu\nu} = \partial_\mu B_\nu - \partial_\nu B_\mu$ is its field strength.
We  define the current,
\begin{equation}
    \mathcal{J}^\mu = (i g_H)\left(H^\dagger \overleftrightarrow{D}^\mu H \right) + \sum_{f}  \left(g^{fL}_{ij} \bar{f}^i_L \gamma^\mu f^j_L+g^{fR}_{ij} \bar{f}^i_R \gamma^\mu f^j_R\right),
 \label{eq:lg2}
\end{equation}
where $H^\dagger \overleftrightarrow{D}^\mu H = H^\dagger (D^\mu H) - (D^\mu H)^\dagger H$ with $D_\mu=\partial_\mu +i g'Y_H B_\mu + i g W^a_{\mu}T^a $ the usual covariant derivative.
The couplings of Eqs.~\ref{eq:lg} and~\eqref{eq:lg2} can be determined for each of the models described in the previous section in terms of the $U(1)^\prime$ gauge coupling $g_D$ and the particle charges of Table~\ref{tab:charges},
\begin{equation}
\label{eq:charges}
g_{H,2}=g_H=Q^H g_D \qquad g_{ij}^{q_L^k}=Q^{q_L^k}g_D\delta_{ij}\qquad g_{ij}^{u_R^k}=Q^{u_R^k}g_D\delta_{ij}\qquad
 g_{ij}^{d_R^k}=Q^{d_R^k}g_D\delta_{ij} .
 \end{equation}
For all of the models except $U(1)_{L_i-L_j}$,
\begin{equation}
g_{ij}^{l_L^k}=Q^{l_L^k}g_D\delta_{ij}\qquad g_{ij}^{e_R^k}=Q^{e_R^k}g_D\delta_{ij}
,
\label{eq:ll}
\end{equation}
while for $U(1)_{L_i-L_j}$ lepton flavour dependence enters,
\begin{equation}
g_{ij}^{l_L^k}=Q^{l_L^k}_{ij}g_D\qquad g_{ij}^{e_R^k}=Q^{e_R^k}_{ij}g_D
.
\label{eq:lll}
\end{equation}

Before any simplifications, after integrating out the heavy $Z'$ we immediately have, up to dimension- 8, 
\begin{align}\label{eq:dim8lagr1}
\begin{split}
\delta \mathcal{L} =& -\frac{1}{2M_{Z'}^2} \mathcal{J}^\mu  \mathcal{J}_\mu + \frac{\epsilon}{M_{Z'}^2}  \left(\partial_{\nu}B^{\mu\nu}\right)\mathcal{J}_\mu - \frac{\epsilon^2}{2M_{Z'}^2} \left(\partial_{\nu}B^{\mu\nu}\right)^2  \\ 
&+\frac{\epsilon^2}{2M_{Z'}^4} \left(\partial_\nu B^{\mu\nu}\right)\partial^2 \left(\partial^\alpha B_{\mu \alpha}\right) - \frac{\epsilon}{M_{Z'}^4}\left(\partial_{\nu}B^{\mu\nu}\right)\partial^2 \mathcal{J}_\mu + \frac{1}{2M_{Z'}^4} \mathcal{J}^\mu \partial^2 \mathcal{J}_\mu\\
&+\frac{\epsilon^2 g_{H,2}^2}{M_{Z'}^4} H^\dagger H \left(\partial_{\nu}B^{\mu\nu}\right)^2 - \frac{2\epsilon g_{H,2}^2}{M_{Z'}^4} H^\dagger H\left(\partial_{\nu}B^{\mu\nu}\right) \mathcal{J}_\mu + \frac{g_{H,2}^2}{M_{Z'}^4} H^\dagger H\mathcal{J}^\mu \mathcal{J}_\mu \, ,
\end{split}
\end{align}
where $\delta \mathcal{L} \equiv \mathcal{L}-\mathcal{L}_\text{SM}$.
We first make a field redefinition to remove the dimension- 6 terms involving $\partial_\nu B^{\mu\nu}$,
\begin{align}
\begin{split}
B_\mu &\rightarrow B_\mu - \frac{\epsilon^2}{2M_{Z'}^2}\left[\left(\partial^\nu B_{\mu\nu}\right) + j_\mu\right] + \frac{\epsilon}{M_{Z'}^2} \mathcal{J}_\mu \, , \\
\partial_\nu B^{\mu\nu} & \rightarrow \partial_\nu B^{\mu\nu} +  \frac{\epsilon^2}{2 M_{Z'}^2} \partial^2 \left(\partial_\nu B^{\mu\nu}\right) + \frac{\epsilon^2}{2M_{Z'}^2} \partial^2 j^\mu - \frac{\epsilon}{M_{Z'}^2}\partial^2 \mathcal{J}^\mu ,
\end{split}
\end{align}
where $j_\mu = -\frac{ig'}{2}\left(H^\dagger \overleftrightarrow{D^\mu} H\right) - g' \sum_f Y_f \bar{f}\gamma^\mu f$ is the SM hypercharge current with $f \in (Q_L,L_L,u_R,d_R,e_R)$,  $g'$ is the SM hypercharge gauge coupling and we omit generation indices for simplicity. 
Note that the definition of $j_\mu$ is the same as in~\cite{Hays:2020scx}.
Including the newly generated dimension- 8 terms, this gives us
\begin{align}
\begin{split}
\delta \mathcal{L} =& -\frac{1}{2M_{Z'}^2} \mathcal{J}^\mu  \mathcal{J}_\mu -\frac{\epsilon^2}{2M_{Z'}^2} j^\mu j_\mu +\frac{\epsilon}{M_{Z'}^2} j^\mu \mathcal{J}_\mu \\ 
&+ \frac{1}{M_{Z'}^4}\bigg\{\left(\frac{\epsilon^2}{2}-\frac{3\epsilon^4}{8}\right)\left(\partial_\nu B^{\mu\nu}\right)\partial^2 \left(\partial^\alpha B_{\mu\alpha}\right) - \left(\epsilon-\epsilon^3\right)\left(\partial_\nu B^{\mu\nu}\right)\partial^2 \mathcal{J}_\mu \\ 
&-\frac{\epsilon^4}{4}\left(\partial_\nu B^{\mu\nu}\right)\partial^2 j_\mu + \left(\frac{1}{2}-\frac{\epsilon^2}{2}\right)\mathcal{J}_\mu \partial^2 \mathcal{J}^\mu+\frac{\epsilon^4}{8}j^\mu \partial^2 j_\mu    - \frac{g'^2 \epsilon^3}{4} \left(H^\dagger H\right)\mathcal{J}^\mu j_\mu \\
& + \left(\epsilon^2 g_{H, 2}^2 + \frac{g'^2 \epsilon^4}{16}\right)\left(H^\dagger H\right) \left(\partial_\nu B^{\mu\nu}\right)^2 - \left(2\epsilon g_{H,2}^2 + \frac{g'^2 \epsilon^3}{4}\right)\left(H^\dagger H \right) \left(\partial_\nu B^{\mu\nu}\right) \mathcal{J}_\mu    \\
& + \left( g_{H,2}^2 + \frac{g'^2 \epsilon^2}{4} \right)\left(H^\dagger H\right) \mathcal{J}^\mu \mathcal{J}_\mu + \frac{g'^2 \epsilon^4}{16}\left(H^\dagger H\right) j_\mu j^\mu + \frac{g'^2 \epsilon^4 }{8}\left(H^\dagger H\right)\left(\partial_\nu B^{\mu\nu }\right) j_\mu
\bigg\} \, .
\end{split}
\end{align}

At dimension- 8, we may use the SM equations of motion to simplify terms since any additional contributions appear at $\mathcal{O}(M_{Z'}^{-6})$.
We have already used this above to eliminate any terms of the form $\partial_\mu j^\mu$ and $\partial_\mu \mathcal{J}^\mu$ for brevity.
Using $\partial_\nu B^{\mu \nu} = j^\mu$ and integration by parts, this yields the simple Lagrangian
\begin{align}\label{eq:Leff}
\begin{split}
\delta \mathcal{L} =& -\frac{1}{2M_{Z'}^2}\left(\mathcal{J}_\mu-\epsilon j_\mu\right)^2 \\ & - \frac{1}{2M_{Z'}^4}\left(1-\epsilon^2\right)\left[\partial_\mu \left(\mathcal{J}_\nu - \epsilon j_\nu\right)\right]^2
+ \frac{1}{M_{Z'}^4}\left(g_{H,2}^2 + \frac{g'^2 \epsilon^2}{4}\right) \left(H^\dagger H\right) \left(\mathcal{J}_\mu - \epsilon j_\mu\right)^2 \, ,
\end{split}
\end{align}
where the first and second lines correspond to the dimension- 6 and dimension- 8 operators, respectively.
We note that we have made no assumptions about the relative size of $\epsilon$.
In the pure kinetic mixing limit, $g_{H,2} = \mathcal{J}_\mu = 0$,  Eq.~\eqref{eq:Leff} clearly matches Eq.~(6.9) of Ref.~\cite{Hays:2020scx}. 

It is worth pointing out that in all of the models we consider, the $Z'$ arises from a UV gauge group, and so the quadratic and linear couplings to the Higgs are forced to be equal by gauge invariance, $g_{H,2}=g_H$. 
This assumption may be relaxed in more complicated scenarios where there is no such constraint, such as composite models.
However, the only place $g_{H,2}$ enters is at dimension-8 in the final term of Eq.~\eqref{eq:Leff}, where it always multiplies another coupling that appears at dimension-6.
The contribution to any observable we will consider from $g_{H,2}$ is therefore always going to be suppressed by a factor $v^2/\Lambda^2$ compared to the leading contribution, making it a small correction even for large values of $g_{H,2}$.
It would be interesting to see how this coupling appears when matching at 1-loop, which we leave for future work.

An alternative approach to matching a generic $Z^\prime$ Lagrangian onto the SMEFT is to transform Eq.~\eqref{eq:lg} into the canonically normalized basis by removing the kinetic mixing term. This is achieved by the field redefinition
\begin{equation}\label{eq:rot}
	\begin{pmatrix} B_\mu\\Z^\prime_\mu\end{pmatrix}=\begin{pmatrix}
		1&\frac{\epsilon}{\sqrt{1-\epsilon^2}}\\0&\frac{1}{\sqrt{1-\epsilon^2}}\end{pmatrix}\begin{pmatrix}
		B_\mu^c\\Z_\mu^{\prime c}	\end{pmatrix}\,,
\end{equation}
where the superscript $c$ denotes gauge fields in the canonical basis. Specifically, this rotation transforms
\begin{equation}
	-\frac14 B_{\mu\nu}B^{\mu\nu}+\frac{\epsilon}{2}B_{\mu\nu}Z^{'\mu\nu}-\frac14 Z'_{\mu\nu}Z^{'\mu\nu}\rightarrow -\frac14 B_{\mu\nu}^cB^{c\mu\nu}-\frac14 Z^{'c}_{\mu\nu}Z^{'c\mu\nu}\,.
\end{equation}
When matching onto SMEFT, this procedure has the advantage that terms involving $\partial_\nu B^{\mu\nu}$ are not present after the $Z'$ has been integrated out, removing the necessity for further field redefinitions. Note, however, that additional contributions arise from the SM Lagrangian, because terms like $(D^{c\mu} H)^\dagger(D^c_\mu H)$ now contain $Z'$ terms from the redefinition of Eq.~\eqref{eq:rot}. Hence, the clear separation between the SM and the new physics Lagrangian is dissolved when eliminating the kinetic mixing.  We have confirmed that both approaches lead to the same matching onto the SMEFT Lagrangian of Eq.~\eqref{eq:Leff}.

The physical mass of the $Z'$ boson is not equivalent to the parameter $M_{Z'}$ in the Lagrangian, Eq.~\eqref{eq:lg}, in the presence of non-vanishing kinetic mixing $\epsilon\neq0$ or coupling to the Higgs $g_H \neq 0$. Transforming to the canonically normalized basis according to Eq.~\eqref{eq:rot} and taking corrections from the neutral $W_3$ and $B$ bosons of the SM weak and hypercharge gauge symmetries into account, we find that the physical masses of the neutral gauge bosons of the SM along with  the $Z'$ boson after electroweak symmetry breaking are given by
\begin{equation}
\label{eq:masses}
	\begin{aligned}
		m_\gamma^2&=0\\
		(m_Z^{\text{phys}})^2&=(m_{Z}^{\text{SM}})^2\left(1-(g^{\prime}\epsilon+2g_DQ_H)^2\frac{v^2}{4M_{Z'}^2}\right)+\mathcal{O}(M_{Z'}^{-4})\\
		(m_{Z'}^{\text{phys}})^2&=\frac{M_{Z'}^2}{(1-\epsilon^2)}+\frac{v^2}{4}\frac{Q_H^2g_D^2+g^{\prime2}\epsilon^2}{1-\epsilon^2}+(g^{\prime}\epsilon+2g_DQ_H)^2\frac{v^2}{4}\frac{(m_Z^\text{SM})^2}{M_{Z'}^2}+\mathcal{O}(M_{Z'}^{-4})\,.
	\end{aligned}
\end{equation}
Here, $	(m_Z^{\text{SM}})^2=\frac{v^2}{4}(g^2+g^{\prime2})$ is the SM (mass)$^2$ of the $Z$. The expansion is valid for large $Z'$ masses and arbitrary mixing parameters $\epsilon$ between the $Z'$ and the hypercharge bosons.
The SM $Z$ couplings and mass are shifted from their SM values due to this mass mixing, an effect that is captured in the SMEFT through the operators $\mathcal{O}_{\phi f}$, $\mathcal{O}_{\phi D}$, as we will see.
We will always work in an expansion in the Lagrangian parameter $M_{Z'}$, though studies have suggested that expanding around the physical mass may lead to improved agreement with the full UV model~\cite{Brehmer:2015rna}.

\subsection{Dimension-6}

While Eq.~\eqref{eq:Leff} is compact, the expressions for the Wilson coefficients are not immediately evident.
Here we include them, starting at dimension-6 in the Warsaw basis~\cite{Grzadkowski:2010es} and using the notation of Ref.~\cite{Dedes:2017zog}.
Inserting the expressions for the currents $\mathcal{J}^\mu \, , j^\mu$ and expanding, we find for $LLLL$ type coefficients
(we omit the superscript $(6)$ in this section),
\begin{eqnarray}
{C_{ll}[ijkl]\over\Lambda^2}&=& -\frac{1}{2M_{Z'}^2} ( g_{ij}^{lL} + \epsilon g' Y_{l} \delta_{ij}) ( g_{kl}^{lL} + \epsilon g' Y_{l} \delta_{kl}) ,\nonumber \\
{C_{lq}^{(1)}[ijkl]\over \Lambda^2}&=& -\frac{1}{M_{Z'}^2} ( g_{ij}^{lL} + \epsilon g' Y_{l} \delta_{ij}) ( g_{kl}^{qL} + \epsilon g' Y_{q} \delta_{kl}) ,\nonumber \\
{C_{qq}^{(1)}[ijkl]\over\Lambda^2}&=&-\frac{1}{2M_{Z'}^2} ( g_{ij}^{qL} + \epsilon g' Y_{q} \delta_{ij}) ( g_{kl}^{qL} + \epsilon g' Y_{q} \delta_{kl}) . 
\end{eqnarray}
Denoting right-handed fermions by $f=u_{R}^k,~d_{R}^k,~e_{R}^k$ with $i,j,k,l$  generation indices, we obtain the $RRRR$ type coefficients,
\begin{eqnarray}
{C_{ff}[ijkl]\over\Lambda^2}&=&-\frac{1}{2M_{Z'}^2} ( g_{ij}^{fR} + \epsilon g' Y_{f} \delta_{ij}) ( g_{kl}^{fR} + \epsilon g' Y_{f} \delta_{kl}) ,\nonumber \\
{C_{ff^\prime}[ijkl]\over\Lambda^2}&=&-\frac{1}{M_{Z'}^2} ( g_{ij}^{fR} + \epsilon g' Y_{f} \delta_{ij}) ( g_{kl}^{f'R} + \epsilon g' Y_{f'} \delta_{kl}) ,\qquad
f\ne f^\prime , \nonumber \\
{C_{ud}^{(1)}[ijkl]\over\Lambda^2}&=&-\frac{1}{M_{Z'}^2} ( g_{ij}^{uR} + \epsilon g' Y_{u} \delta_{ij}) ( g_{kl}^{dR} + \epsilon g' Y_{d} \delta_{kl}).
\end{eqnarray}
The mixed $LLRR$ $\psi^4$ coefficients are 
\begin{eqnarray}
{C_{lf}[ijkl]\over\Lambda^2}&=& -\frac{1}{M_{Z'}^2} ( g_{ij}^{lL} + \epsilon g' Y_{l} \delta_{ij}) ( g_{kl}^{fR} + \epsilon g' Y_{f} \delta_{kl}) ,\nonumber \\
{C_{qf}^{(1)}[ijkl]\over\Lambda^2}&=& -\frac{1}{M_{Z'}^2} ( g_{ij}^{qL} + \epsilon g' Y_{q} \delta_{ij}) ( g_{kl}^{fR} + \epsilon g' Y_{f} \delta_{kl}).
\end{eqnarray}
In the operator class $\psi^2 H^2 D$, we find the coefficients,
\begin{align}
    \frac{C_{\varphi l}^{(1)}[ij]}{\Lambda^2} &=  -\frac{1}{2M_{Z'}^2}(2g_H+\epsilon g') ( g_{ij}^{lL} + \epsilon g' Y_{l} \delta_{ij}), \\
    \frac{C_{\varphi q}^{(1)}[ij]}{\Lambda^2} &=  -\frac{1}{2M_{Z'}^2}(2g_H+\epsilon g') ( g_{ij}^{qL} + \epsilon g' Y_{q} \delta_{ij}), \\
    \frac{C_{\varphi f}[ij]}{\Lambda^2} &=  -\frac{1}{2M_{Z'}^2}(2g_H+\epsilon g') ( g_{ij}^{fL} + \epsilon g' Y_{f} \delta_{ij}).
\end{align}
Finally, we obtain the coefficient, 
\begin{equation}\label{eq:OTdef}
\frac{1}{8M_{Z'}^2}\left(2g_H+\epsilon g'\right)^2 (H^\dagger \overleftrightarrow{D}_\mu H)^2 = \frac{1}{4M_{Z'}^2}\left(2g_H+\epsilon g'\right)^2\mathcal{O}_T,
\end{equation}
where $\mathcal{O}_T \equiv \frac{1}{2}(H^\dagger \overleftrightarrow{D}_\mu H)^2$ is the operator that generates the oblique $T$ parameter.
This may be decomposed into a combination of $\mathcal{O}_{\varphi \square}$ and $\mathcal{O}_{\varphi D}$ via the relation $\mathcal{O}_T = -\frac{1}{2}\mathcal{O}_{\varphi\square} - 2\mathcal{O}_{\varphi D}$ to yield
\begin{align}
\frac{C_{\varphi \square}}{\Lambda^2} &= -\frac{1}{8M_{Z'}^2}\left(2g_H+\epsilon g'\right)^2,\\
\frac{C_{\varphi D}}{\Lambda^2} &= -\frac{1}{2M_{Z'}^2}\left(2g_H+\epsilon g' \right)^2.
\end{align}
Note that since $T\propto C_T\propto \left(2g_H+\epsilon g'\right)^2$, the contribution to the oblique $T$ parameter is always positive.
In general, we see that the presence of an $\epsilon$ contribution allows for many directions with vanishing 4-fermion coefficients.

\subsection{Dimension-8}
\setlength{\tabcolsep}{0.5em} 
{\renewcommand{\arraystretch}{1.2}
\begin{table}[t]
\centering
\begin{tabular}{||c|c||c|c||}
\hline
\hline

$\mathcal{O}^{(1)}_{l^4 H^2}[ijkl] $ & $(\bar{l}_i\gamma^\mu l_j)(\bar{l}_k\gamma_\mu l_l) (H^\dagger H)$
& $\mathcal{O}^{(1)}_{l^4 D^2}[ijkl] $ & $D^\nu(\bar{l}_i\gamma^\mu l_j)D_\nu(\bar{l}_k\gamma_\mu l_l)$ \\
$\mathcal{O}^{(1)}_{q^4 H^2}[ijkl] $ & $(\bar{q}_i\gamma^\mu q_j)(\bar{q}_k\gamma_\mu q_l) (H^\dagger H)$
& $\mathcal{O}^{(1)}_{q^4 D^2}[ijkl] $ & $D^\nu (\bar{q}_i\gamma^\mu q_j)D_\nu(\bar{q}_k\gamma_\mu q_l)$ \\
$\mathcal{O}^{(1)}_{l^2 q^2 H^2}[ijkl] $ & $(\bar{l}_i\gamma^\mu l_j)(\bar{q}_k\gamma_\mu q_l) (H^\dagger H)$
& $\mathcal{O}^{(1)}_{l^2 q^2 D^2}[ijkl] $ & $D^\nu(\bar{l}_i\gamma^\mu l_j)D_\nu(\bar{q}_k\gamma_\mu q_l)$ \\
\hline 

$\mathcal{O}_{f^4 H^2}[ijkl] $ & $(\bar{f}_i\gamma^\mu f_j)(\bar{f}_k\gamma_\mu f_l) (H^\dagger H)$
& $\mathcal{O}_{f^4 D^2}[ijkl] $ & $D^\nu(\bar{f}_i\gamma^\mu f_j)D_\nu(\bar{f}_k\gamma_\mu f_l)$\\
$\mathcal{O}_{f^2 f'^2 H^2}[ijkl] $ & $(\bar{f}_i\gamma^\mu f_j)(\bar{f'}_k\gamma_\mu f'_l) (H^\dagger H)$
& $\mathcal{O}_{f^2 f'^2 D^2}[ijkl] $ & $D^\nu(\bar{f}_i\gamma^\mu f_j)D_\nu(\bar{f'}_k\gamma_\mu f'_l)$ \\
$\mathcal{O}^{(1)}_{u^2 d^2 H^2}[ijkl] $ & $(\bar{u}_i\gamma^\mu u_j)(\bar{d}_k\gamma_\mu d_l) (H^\dagger H)$
& $\mathcal{O}^{(1)}_{u^2 d^2 D^2}[ijkl] $ & $D^\nu (\bar{u}_i\gamma^\mu u_j) D_\nu(\bar{d}_k\gamma_\mu d_l)$ \\
\hline

$\mathcal{O}^{(1)}_{l^2 f^2 H^2}[ijkl] $ & $(\bar{l}_i\gamma^\mu l_j)(\bar{f}_k\gamma_\mu f_l) (H^\dagger H)$
& $\mathcal{O}^{(1)}_{l^2 f^2 D^2}[ijkl] $ & $D^\nu(\bar{l}_i\gamma^\mu l_j)D_\nu(\bar{f}_k\gamma_\mu f_l)$ \\

$\mathcal{O}^{(1)}_{q^2 f^2 H^2}[ijkl] $ & $(\bar{q}_i\gamma^\mu q_j)(\bar{f}_k\gamma_\mu f_l) (H^\dagger H)$
& $\mathcal{O}^{(1)}_{q^2 f^2 D^2}[ijkl] $ & $D^\nu(\bar{q}_i\gamma^\mu q_j)D_\nu(\bar{f}_k\gamma_\mu f_l)$ \\
\hline

$\mathcal{O}_{f^2 H^4 D}[ij]$  &  $i(\bar{f}_{i}\gamma^\mu f_j)(H^\dagger \overleftrightarrow{D}_\mu H) (H^\dagger H) $ 
& $\mathcal{O}^{(8)}_{D^2 \phi f}[ij]$ & $iD_\nu (\bar{f}_{i}\gamma^\mu f_j)D^\nu(H^\dagger \overleftrightarrow{D}_\mu H) $\\
$\mathcal{O}^{(1)}_{l^2 H^4 D}[ij]$ & $i(\bar{l}_{i}\gamma^\mu l_j)(H^\dagger \overleftrightarrow{D}_\mu H) (H^\dagger H) $ 
& $\mathcal{O}^{(8)}_{D^2 \phi l}[ij]$ & $iD_\nu (\bar{l}_{i}\gamma^\mu l_j)D^\nu(H^\dagger \overleftrightarrow{D}_\mu H) $ \\
$\mathcal{O}^{(1)}_{q^2 H^4 D}[ij] $ & $i(\bar{q}_{i}\gamma^\mu q_j)(H^\dagger \overleftrightarrow{D}_\mu H) (H^\dagger H) $ 
& $\mathcal{O}^{(8)}_{D^2 \phi q}[ij]$ & $iD_\nu (\bar{q}_{i}\gamma^\mu q_j)D^\nu(H^\dagger \overleftrightarrow{D}_\mu H) $\\
\hline

$\mathcal{O}_{T}^{(8)}$ & $\frac{1}{2}(H^\dagger H) (H^\dagger \overleftrightarrow{D}_\mu H)^2$ & $\mathcal{O}_{D^4\phi^4}^{(8)}$ 
& $\frac{1}{2}(H^\dagger \overleftrightarrow{D}_\mu H)\square (H^\dagger \overleftrightarrow{D}^\mu H)$ \\

$\mathcal{O}_{\phi \square}^{(8)}$ & $(H^\dagger H)^2 \square (H^\dagger H)$ & $\mathcal{O}_{\phi D}^{(8)}$ & $(H^\dagger H)|H^\dagger D_\mu H|^2$ \\

\hline
\hline
\end{tabular}
\caption{Definitions for the dimension-8 operators generated in this paper. Here $f^k=(e_{R}^k,u_{R}^k,d_{R}^k)$ denotes right handed fermions with generation index $k$, and $f\neq f'$. We use the notation of Ref.~\cite{Murphy:2020rsh} for operators already in that basis, while others are indicated by the superscript $(8)$.
\label{tab:dim8ops}}
\end{table}
}

Since all operators come from terms of the form $(\mathcal{J}_\mu - \epsilon j_\mu)^2$, the operators at dimension-8 may be constructed in terms of the dimension-6 coefficients of the previous section in an obvious way.
We define our notation for the generated operators in Table~\ref{tab:dim8ops}, using the notation of Ref.~\cite{Murphy:2020rsh} for operators corresponding to that basis and a superscript $(8)$ for the remainder of the operators.
To keep the connection to the dimension-6 operators manifest, we do not attempt to fully translate into the basis of Ref.~\cite{Murphy:2020rsh}.
We suppress flavour indices throughout since they are the same between the dimension-8 and dimension-6 pieces.
Starting with the four-fermion operators with $H^\dagger H$ factors, we find
\begin{equation}
\begin{split}
\frac{C_{l^4 H^2}^{(1)}}{\Lambda^4}&= -\left(\frac{4g_{H,2}^2 + g'^2 \epsilon^2}{2 M_{Z'}^2}\right) \frac{C_{ll}}{\Lambda^2}, \\
\frac{C_{q^4 H^2}^{(1)}}{\Lambda^4}&= -\left(\frac{4g_{H,2}^2 + g'^2 \epsilon^2}{2 M_{Z'}^2}\right) \frac{C_{qq}^{(1)}}{\Lambda^2}, \\
\frac{C_{f^2 f'^2 H^2}}{\Lambda^4}&= -\left(\frac{4g_{H,2}^2 + g'^2 \epsilon^2}{2 M_{Z'}^2}\right) \frac{C_{ff'}}{\Lambda^2}, \\
\frac{C_{l^2 f^2 H^2}^{(1)}}{\Lambda^4}&= -\left(\frac{4g_{H,2}^2 + g'^2 \epsilon^2}{2 M_{Z'}^2}\right)\frac{C_{lf}^{(1)}}{\Lambda^2}, 
\end{split}
\qquad
\begin{split}
\frac{C_{l^2 q^2 H^2}^{(1)}}{\Lambda^4}&= -\left(\frac{4g_{H,2}^2 + g'^2 \epsilon^2}{2 M_{Z'}^2}\right)\frac{C_{lq}^{(1)}}{\Lambda^2} ,\\
\frac{C_{f^4 H^2}}{\Lambda^4}&= -\left(\frac{4g_{H,2}^2 + g'^2 \epsilon^2}{2 M_{Z'}^2}\right) \frac{C_{ff}}{\Lambda^2},\\
\frac{C_{u^2 d^2 H^2}^{(1)}}{\Lambda^4}&= -\left(\frac{4g_{H,2}^2 + g'^2 \epsilon^2}{2 M_{Z'}^2}\right)\frac{C_{ud}^{(1)}}{\Lambda^2},\\
\frac{C_{q^2 f^2 H^2}^{(1)}}{\Lambda^4}&= -\left(\frac{4g_{H,2}^2 + g'^2 \epsilon^2}{2 M_{Z'}^2}\right)\frac{C_{qf}^{(1)}}{\Lambda^2}\, .
\end{split}
\end{equation}
Similarly, for the four-fermion operators with two derivatives, we have the matched coefficients,
\begin{equation}
\begin{split}
\frac{C_{l^4 D^2}^{(1)}}{\Lambda^4} &= \left(\frac{1-\epsilon^2}{M_{Z'}^2}\right) \frac{C_{ll}}{\Lambda^2}, \\
\frac{C_{q^4 D^2}^{(1)}}{\Lambda^4} &= \left(\frac{1-\epsilon^2}{M_{Z'}^2}\right) \frac{C_{qq}^{(1)}}{\Lambda^2}, \\
\frac{C_{f^2 f'^2 D^2}}{\Lambda^4} &= \left(\frac{1-\epsilon^2}{M_{Z'}^2}\right) \frac{C_{f f'}}{\Lambda^2}, \\
\frac{C_{l^2 f^2 D^2}^{(1)}}{\Lambda^4} &= \left(\frac{1-\epsilon^2}{M_{Z'}^2}\right)\frac{C_{lf}^{(1)}}{\Lambda^2}, 
\end{split}
\qquad 
\begin{split}
\frac{C_{l^2 q^2 D^2}^{(1)}}{\Lambda^4} &= \left(\frac{1-\epsilon^2}{M_{Z'}^2}\right) \frac{C_{lq}^{(1)}}{\Lambda^2}, \\
\frac{C_{f^4 D^2}}{\Lambda^4} &= \left(\frac{1-\epsilon^2}{M_{Z'}^2}\right) \frac{C_{ff}}{\Lambda^2}, \\
\frac{C_{u^2 d^2 D^2}^{(1)}}{\Lambda^4} &= \left(\frac{1-\epsilon^2}{M_{Z'}^2}\right) \frac{C_{ud}^{(1)}}{\Lambda^2}, \\
\frac{C_{q^2 d^2 D^2}^{(1)}}{\Lambda^4} &= \left(\frac{1-\epsilon^2}{M_{Z'}^2}\right) \frac{C_{qf}^{(1)}}{\Lambda^2}.
\end{split}
\end{equation}
Moving on to two-fermion operators, we find the following coefficients:
\begin{equation}
\begin{split}
\frac{C_{f^2 H^4 D}}{\Lambda^4}&= -\left(\frac{4g_{H,2}^2 + g'^2 \epsilon^2}{2 M_{Z'}^2}\right) \frac{C_{\phi f}}{\Lambda^2}, \\
\frac{C_{l^2 H^4 D}^{(1)}}{\Lambda^4}&= -\left(\frac{4g_{H,2}^2 + g'^2 \epsilon^2}{2 M_{Z'}^2}\right) \frac{C_{\phi l}^{(1)}}{\Lambda^2}\\
\frac{C_{q^2 H^4 D}^{(1)}}{\Lambda^4}&= -\left(\frac{4g_{H,2}^2 + g'^2 \epsilon^2}{2 M_{Z'}^2}\right)\frac{C_{\phi q}^{(1)}}{\Lambda^2}, 
\end{split}
\qquad
\begin{split}
\frac{C_{D^2 \phi f}^{(8)}}{\Lambda^4}&= \left(\frac{1-\epsilon^2}{M_{Z'}^2}\right)\frac{C_{\phi f}}{\Lambda^2}, \\
\frac{C_{D^2 \phi l}^{(8)}}{\Lambda^4}&= \left(\frac{1-\epsilon^2}{M_{Z'}^2}\right)\frac{C_{\phi l}^{(1)}}{\Lambda^2}, \\
\frac{C_{D^2 \phi q}^{(8)}}{\Lambda^4}&= \left(\frac{1-\epsilon^2}{M_{Z'}^2}\right)\frac{C_{\phi q}^{(1)}}{\Lambda^2}\, ,
\end{split}
\end{equation}
where we should emphasize that the operators with a superscript $(8)$ are not 
written in the basis of Ref.~\cite{Murphy:2020rsh}.
Finally, there are two bosonic operators that are generated, with coefficients given by
\begin{align}
\begin{split}
\frac{C_{T}^{(8)}}{\Lambda^4}&= -\left(\frac{4g_{H,2}^2 + g'^2 \epsilon^2}{2 M_{Z'}^2}\right) \frac{C_{T}}{\Lambda^2}, \\
\frac{C_{D^4 \phi^4}^{(8)}}{\Lambda^4}&= \left(\frac{1-\epsilon^2}{M_{Z'}^2}\right)\frac{C_{T}}{\Lambda^2}, \\
\end{split}
\end{align}
where $C_T$ is defined by Eq.~\eqref{eq:OTdef}.
Of these two operators, only $\mathcal{O}_{T}^{(8)}$ contributes to the observables we will consider. 
In terms of the operators in Table~\ref{tab:dim8ops}, $\mathcal{O}_T^{(8)}$ is redundant due  to a similar relation to that we used for $\mathcal{O}_T$ at dimension-6.
In particular, we have
\begin{align}
\begin{split}
\mathcal{O}_T^{(8)} &= \frac{1}{2}(H^\dagger H)(H^\dagger \overleftrightarrow{D}_\mu H)^2 \, ,\\
&= -\frac{1}{4} (H^\dagger H)^2 \square (H^\dagger H) - 2 (H^\dagger H)|H^\dagger D_\mu H|^2 \, , \\
&\equiv -\frac{1}{4}\mathcal{O}_{\phi \square}^{(8)} - 2\mathcal{O}_{\phi D}^{(8)}\, .
\end{split}
\label{eq:red}
\end{align}
Since the relationship at dimension-6 is identical to that of Eq.~\eqref{eq:red} other than a factor of 2, we have 
\begin{align}
\begin{split}
\frac{C_{\phi \square}^{(8)}}{\Lambda^4} &= -\left(\frac{4g_{H,2}^2 + g'^2 \epsilon^2}{4 M_{Z'}^2}\right) \frac{C_{\phi \square}}{\Lambda^2}, \\
\frac{C_{\phi D}^{(8)}}{\Lambda^4} &= -\left(\frac{4g_{H,2}^2 + g'^2 \epsilon^2}{2M_{Z'}^2}\right) \frac{C_{\phi D}}{\Lambda^2}.
\end{split}
\end{align}

\section{Results}
\label{sec:res}
We are now in a position to present our numerical results and compare constraints from electroweak precision observables (EWPOs) and invariant mass measurements and forward backward asymmetries from neutral Drell Yan production at the LHC.  For each data set, we obtain limits for the  matching scenarios presented in the previous section corresponding to  the UV complete models of Section \ref{sec:zp}.  SMEFT observables are expanded as
\begin{equation}
d\sigma = d\sigma^\text{SM} + \frac{1}{\Lambda^2} \sum_i a^{(6)}_i C^{(6)}_i + \frac{1}{\Lambda^4} \left(\sum_{ij} b^{(6)}_{ij} C^{(6)}_i C^{(6)}_j + \sum_i a^{(8)}_i C^{(8)}_i \right) \, , 
\end{equation}
where the numerical factors,  $a^{(6)}_i,  b^{(6)}_{ij} ,$ and  $a^{(8)}_i$ are process dependent.
 
Bounds on SMEFT coefficients can be derived from fits to the EWPOs~\cite{Dawson:2022bxd} and in many UV models, these provide the most stringent constraints.  The $Z$ and $W$ boson pole observables that we consider are
\begin{eqnarray}
&&M_W,\, \Gamma_W, \,\Gamma_Z, \, \sigma_h,\,  R_e, \, R_\mu,\, R_\tau, \, R_c,\, R_b,\, A_e,\nonumber \\ &&  A_\mu,\, A_\tau, \,A_c,\, A_b,\, A_{e,\text{FB}},\,A_{\mu,\text{FB}},\,A_{\tau,\text{FB}}, \, A_{c,\text{FB}},\,A_{b,\text{FB}} .
\label{eq:ewpo}
\end{eqnarray} 
We perform a $\chi^2$ fit to the data in Table III of Ref.~\cite{Bellafronte:2023amz}, using as the SM contribution the most precisely known theoretical values given in this table.  Both the (dimension-6)$^2$ contributions (the $b^{(6)}_{ij}/\Lambda^4$ terms)
 and the dimension-8 contributions (the $a^{(8)}_i/\Lambda^4$ terms)  are included at tree level for each $Z^\prime$ model described in the previous section.  Following Ref.~\cite{Bellafronte:2023amz}, we allow for an arbitrary flavor structure in the leptonic sector and  take as our   input parameters,
$G_\mu=1.1663787(6)\times 10^{-5}~\text{GeV}^{-2},~
m^\text{phys}_Z=91.1876\pm .0021~\text{GeV},~
m_W^\text{phys}=80.379\pm 0.012~\text{GeV},~
\alpha_s(m^\text{phys}_Z)=0.1181\pm 0.0011,~
M_h=125.25\pm 0.17~\text{GeV}, $ and $
M_t=172.69\pm 0.5~\text{GeV}$.

In addition to EWPOs, we include high invariant mass Drell-Yan data from four LHC datasets in our SMEFT fits.  Drell-Yan (DY) is a sensitive probe of 4-fermion interactions since at dimension-6, these interactions are
enhanced by a factor of (energy)$^2$ relative to the SM tree level result~\cite{deBlas:2013qqa,Dawson:2021ofa,Boughezal:2022nof,Allwicher:2022gkm,Allwicher:2022mcg}.  This is in contrast to the EWPOs results, where SMEFT effects scale as $(m_Z^\text{phys})^2/\Lambda^2$.
The SMEFT limits on $Z^\prime$ models are often interpreted in terms of oblique parameters~\cite{Farina:2016rws,Panico:2021vav,Torre:2020aiz}, although the complete SMEFT fit can yield additional information. 
The DY data we include consists of $d\sigma/dm_{\ell\ell}$ measurements at 8 TeV~\cite{ATLAS:2016gic} and 13 TeV~\cite{CMS:2021ctt}, as well as high invariant mass forward-backward asymmetry ($A_\text{FB}$) measurements at 8 TeV~\cite{CMS:2016bil} and 13 TeV~\cite{CMS:2022uul}\footnote{
    We assume these datasets to be uncorrelated. Note that any non-zero correlation should have minimal impact on our final results since they are driven primarily by the 13 TeV $d\sigma /dm_{\ell\ell}$ distributions.
}.
These datasets were recently considered in Ref.~\cite{Boughezal:2023nhe} as well.
Note that while we will refer to both measurements simply as $A_\text{FB}$, the 8 TeV and 13 TeV CMS studies use different definitions. 
The forward backward asymmetry $A_\text{FB}$ is defined by
\begin{equation}
A_\text{FB} \equiv \frac{\Delta \sigma}{\sigma} = \frac{\sigma_\text{F}-\sigma_\text{B}}{\sigma_\text{F}+\sigma_\text{B}} \, ,
\end{equation}
where $\sigma_\text{F}$ and $\sigma_\text{B}$ are the total cross sections for forward and backward events, defined by $\cos{\theta} >0$ and $\cos{\theta} < 0$, respectively, where $\theta$ is the angle between the incoming quark and outgoing negatively charged lepton in the dilepton center of mass frame.
In the lab frame, this angle is given by
\begin{equation}
\cos{\theta^*} = \frac{2(P_1^+ P_2^- - P_1^- P_2^+)}{\sqrt{K^2(K^2+K_T^2)}} \, ,
\end{equation}
where $P_i^\pm = (E_i \pm p_i^z)/\sqrt{2}$, with $E_i$ and $p_i$ the energy and four-momentum of lepton $i=1(2)$ for $\ell^-(\ell^+)$, and $K=p_1+p_2$.
However, to define the positive axis, this definition requires the identification of the quark direction, which is only accessible in simulation.
In the 8 TeV study, CMS approximately identifies this with the $z$-component of the dilepton invariant mass,
\begin{equation}
\cos \theta = \frac{|K^z|}{K^z}\cos \theta^* \, ,
\end{equation}
while in the 13 TeV study, CMS identifies it with a template fit using Monte Carlo simulations where one has access to truth level information.

Owing to this complication, we use the SM predictions and associated uncertainties provided by CMS for the 13 TeV $A_\text{FB}$ measurement. 
For our SMEFT predictions, we will manually pick out the truth-level quark direction from our Monte-Carlo as an approximation for their procedure, which was seen in Ref.~\cite{Boughezal:2023nhe} to yield reasonable agreement with CMS for the SM prediction.
Likewise, for the 13 TeV $d\sigma/dm_{\ell\ell}$ measurement, the results are not unfolded by CMS, and so we cannot directly compare with our own predictions without background estimates for the non-DY backgrounds. 
We use the CMS SM predictions for this dataset as well.
For the 8 TeV $d\sigma/dm_{\ell\ell}$ and $A_\text{FB}$ measurements~\cite{ATLAS:2016gic}, we compute the SM predictions to NLO QCD using MCFM~\cite{Campbell:2019dru} with NNPDF 3.1 parton distribution functions~\cite{NNPDF:2017mvq}\footnote{We find small differences of $\mathcal{O}(3-5\%)$ in our results when using NNPDF3.0
parton distribution functions.}.
While the NNLO SM QCD corrections are also available, we have checked that their impact is numerically small, so we neglect them.
Electroweak Sudakov logarithms become important at high invariant masses and must also be included. 
We use the exact electroweak corrections as computed by MCFM, multiplying our SM NLO QCD distributions by $k$-factors defined bin-by-bin as $k_i = (\sigma^i_\text{LO}+\delta^i_\text{EW}) / \sigma^i_\text{LO}$ for bin $i$.
For the invariant mass distributions and the 8 TeV $A_\text{FB}$ dataset we use a scale $\mu_{R}=\mu_{F}=m_{\ell\ell}$, while for the 13 TeV $A_\text{FB}$ we use a scale $\mu_{R}=\mu_{F}=H_T$ following Ref.~\cite{Boughezal:2023nhe}, where $H_T$ is the sum of the final state transverse momenta. 
Our theory uncertainties for the 8 TeV datasets are computed by a 6-point scale variation around the central scale value $\mu_{R,F} = (1/2m_{\ell\ell},m_{\ell\ell},2m_{\ell\ell})$ and turning the maximum deviation for each bin into a symmetric uncertainty.

We compute the SMEFT contributions for the DY observables at LO QCD using MadGraph5~\cite{Alwall:2014hca}. 
We obtain our SMEFT model files using a combination of SMEFTsim~\cite{Brivio:2020onw} and SmeftFR~\cite{Dedes:2023zws}, modified to add the relevant fermionic dimension-8 operators.
The forward backward asymmetries are computed to ${\cal{O}}({1/\Lambda^4})$, 
\begin{align}\label{eq:AFBsmeft}
\begin{split}
A_\text{FB} =A_\text{FB}^\text{SM} &+ \frac{1}{\Lambda^2}\sum_i C^{(6)}_i \left(\frac{\Delta {\hat{a}}_i^{(6)}}{\sigma_\text{SM}}-\frac{{\hat{a}}_i^{(6)}\Delta \sigma_\text{SM}}{\sigma_\text{SM}^2}\right) \\
& + \frac{1}{\Lambda^4}\sum_{ij} C^{(6)}_i C^{(6)}_j \left[\frac{\Delta {\hat {b}}^{(6)}_{ij}}{\sigma_\text{SM}}
-\left(\frac{{\hat{b}}^{(6)}_{ij}\Delta \sigma_\text{SM} +{\hat{a}}_i^{(6)}\Delta {\hat{a}}_j^{(6)}}{\sigma^2_\text{SM}}\right) + \frac{{\hat{a}}_i^{(6)} {\hat{a}}_j^{(6)}\Delta \sigma_\text{SM}}{\sigma_\text{SM}^3}\right] \\ &+\frac{1}{\Lambda^4}\sum_i C^{(8)}_i \left(\frac{\sigma_\text{SM} \Delta {\hat{a}}_i^{(8)} - {\hat{a}}_i^{(8)}\Delta \sigma_\text{SM}}{\sigma_\text{SM}^2} \right)\, ,
\end{split}
\end{align}
where $\Delta\sigma_\text{SM}$ is computed as in the previous paragraph, $\Delta (a,b)$ are defined analoguously to $\Delta \sigma$,
and we note that the coefficients ${\hat{a}}_i^{(6)},~ {\hat {b}}^{(6)}_{ij}$ and ${\hat{a}}_i^{(8)}$ are specific to the DY process and depend on the energy and experimental cuts . 
For the $\sigma_\text{SM}$ and $\Delta\sigma_\text{SM}$ appearing in the SMEFT expansion of Eq.~\eqref{eq:AFBsmeft} we use LO predictions to match the order of our SMEFT predictions, while for $A_\text{FB}^\text{SM}$ we use the full NLO+EW MCFM or CMS predictions.

\begin{figure}
    \centering
    \includegraphics[width=0.9\textwidth]{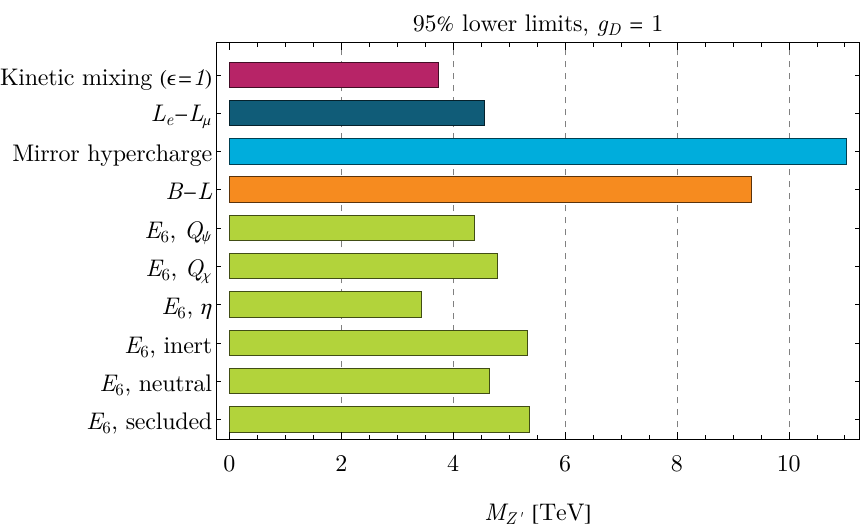}
    \caption{The 95\% CL exclusion limit on $M_{Z'}$ for our considered models assuming the associated coupling $g_D=1$ and kinetic mixing $\epsilon=0$. For the case of pure kinetic mixing we set $\epsilon=1$ instead. Note that $M_{Z'}$ is smaller than the physical mass $m^\text{phys}_{Z'}$ when there is a non-zero $\epsilon$ or $Q_H$.}
    \label{fig:Mmin}
\end{figure}

\begin{figure}
    \centering
    \includegraphics[width=0.9\textwidth]{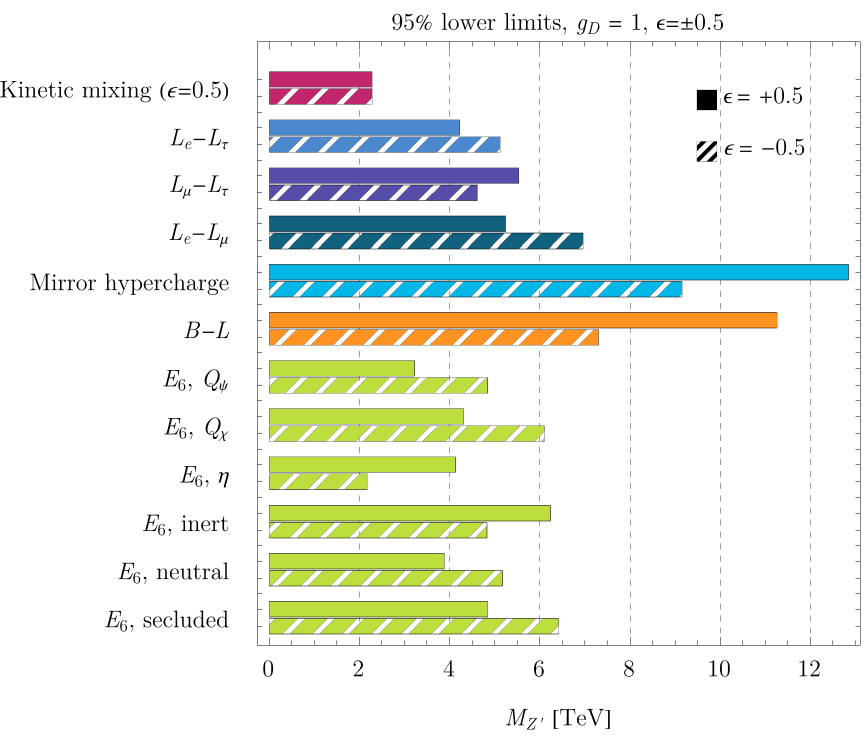}
    \caption{The 95\% CL exclusion limit on $M_{Z'}$ for our considered models assuming the associated coupling $g_D=1$ and kinetic mixing $\epsilon=\pm 0.5$, where the solid (hatched) bars have positive (negative) $\epsilon$. Note that $M_{Z'}$ is smaller than the physical mass $m^\text{phys}_{Z'}$ when there is a non-zero $\epsilon$ or $Q_H$. With $|\epsilon| = 0.5$, $m^\text{phys}_{Z'} \approx 1.15 \times M_{Z'}$ for the models shown here.}
    \label{fig:MminEpPt5}
\end{figure}

Other than the  pure kinetic mixing model, all of the models discussed in Section~\ref{sec:zp} are three-parameter models, corresponding to the mass $M_{Z'}$, the gauge coupling $g_D$, and kinetic mixing $\epsilon$. 
To begin, we first show the 95\% lower limits on $M_{Z'}$ in Fig.~\ref{fig:Mmin}, where we set $\epsilon=1$ for the pure kinetic mixing case, and $g_D=1$ with $\epsilon=0$ for other models. 
This includes the combination of EWPOs, and  $A_\text{FB}$, and $d\sigma/dm_{\ell\ell}$ from our DY data sets, with all SMEFT contributions computed up to ${\cal{O}}({1/\Lambda^4})$.
The mirror hypercharge and gauged $B-L$ models are strongly constrained to $M_{Z'} \gtrsim 10$ TeV,  since they both generate the semileptonic four-fermion operators relevant for Drell-Yan. 
In the pure kinetic mixing case $(\epsilon=1)$, the lower constraining power compared to the mirror hypercharge model is entirely from $g' \sim 0.3$, (as opposed to $g_D=1$ shown in the figures), since the pattern of operators is the same. 
Likewise, the small fractional charges in the $E_6$ models lead to a significantly weaker mass reach compared to the mirror hypercharge and $B-L$ models.
The $L_e-L_\mu$ model does not generate any semileptonic four-fermion operators, and so the constraining power is much weaker.
We do not show the $L_\mu-L_\tau$ or $L_e-L_\tau$ models in Fig.~\ref{fig:Mmin}, as they are completely unconstrained by the observables we consider when $\epsilon=0$.
In Fig.~\ref{fig:MminEpPt5}, we set $\epsilon=\pm 0.5$ to show the impact of kinetic mixing on the $M_{Z'}$ exclusion limits. 
In this case, constraints on the $L_\mu-L_\tau$ and $L_e-L_\tau$ models emerge at the 4-5 TeV level, and the constraints on the other models either strengthen or weaken significantly depending on the sign of $\epsilon$. 

These limits on $M_{Z'}$ can be compared to those obtained through general SMEFT fits derived in the literature using
EWPOs, Higgs data, diboson production and top data~\cite{Bartocci:2023nvp,Ellis:2018gqa,Dawson:2020oco,Grojean:2018dqj,daSilvaAlmeida:2018iqo,Biekotter:2018ohn,delAguila:2011zs}.
In a global fit, all SMEFT operators of the Warsaw basis are taken into account, and the new physics model is unspecified. 
These fits can be extended by assuming that the coefficients have the pattern corresponding to a single new heavy particle  
with general couplings to the SM particles~\cite{terHoeve:2023pvs,delAguila:2010mx,Ellis:2018gqa,deBlas:2017xtg}. 
Performing a global fit with these restrictions  on the Wilson coefficients, it is possible to derive a lower mass limit for the $Z'$,
where the limits come almost exclusively from EWPOs.   
One typically finds $M_{Z^\prime}^\text{global}\gtrsim\mathcal{O}(1-\text{few})$~TeV. 
We are able to derive slightly stronger constraints, i.e. higher lower mass limits, than these  SMEFT fits, because in our framework where specific $Z'$ models are examined, the couplings to the $Z^\prime$ are not arbitrary, but are related as in Eqs. \eqref{eq:charges}-\eqref{eq:lll} and also because of the inclusion of the DY contributions which has a significant impact. Since most of our  $Z'$ models still induce a plethora of SMEFT operators, our constraints are of the same order of magnitude as the fits derived with generic couplings to a heavy neutral gauge boson~\cite{terHoeve:2023pvs,delAguila:2010mx}.
Higher $M_{Z'}$ constraints derived from individual operator fits where only one SMEFT Wilson coefficient is assumed to be non-vanishing at a time are to be taken with caution, because a realistic new physics scenario is very unlikely to generate only one non-vanishing SMEFT operator.

\begin{figure}
    \centering
    \includegraphics[width=0.495\textwidth]{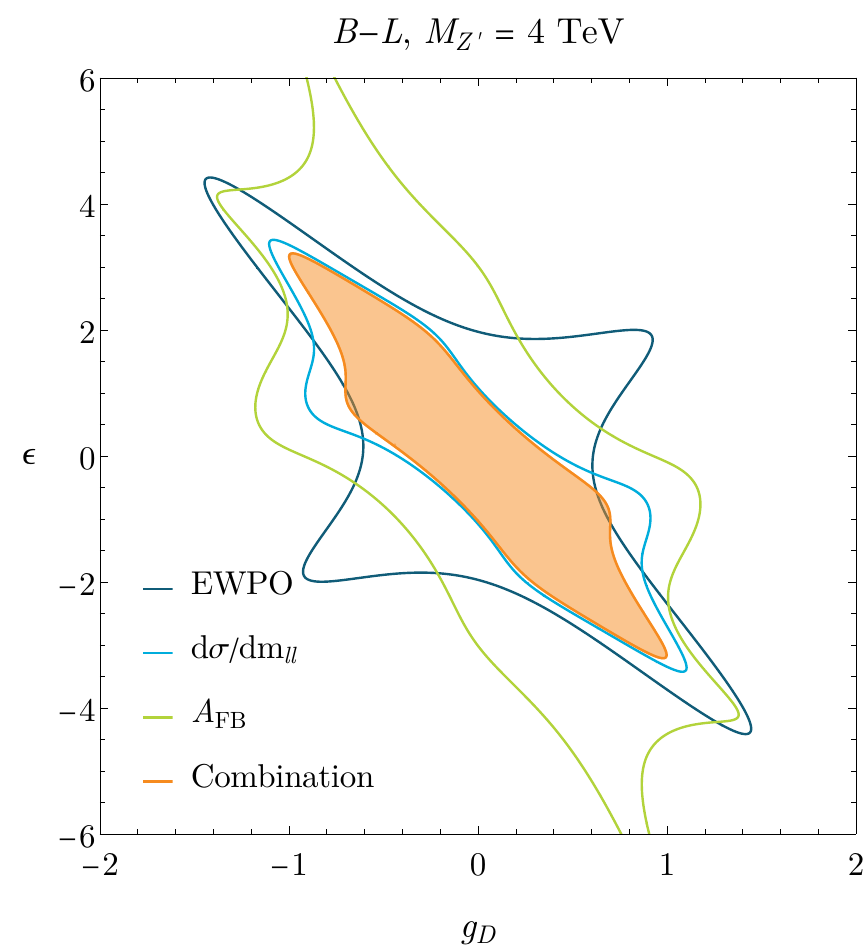}
    \includegraphics[width=0.495\textwidth]{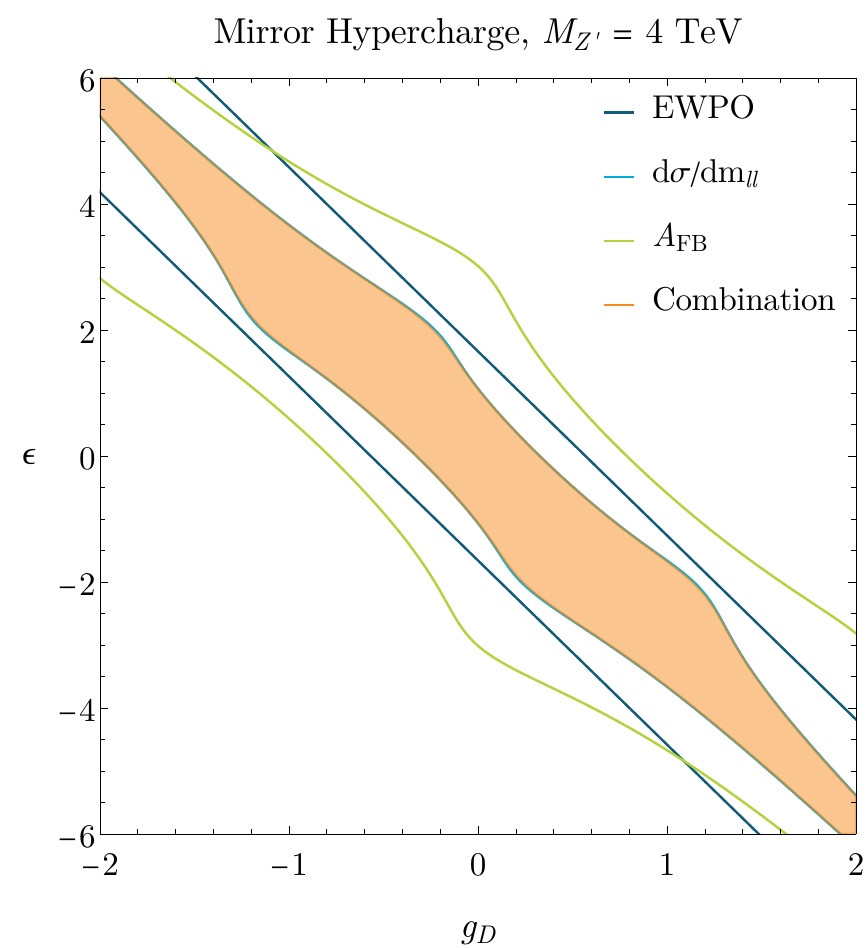}
    \caption{95\% CL constraints in the $(g_D,\epsilon)$ plane for the $B-L$ and mirror hypercharge models with a benchmark 
    mass of $M_{Z^\prime}=4$ TeV. We show both separately and in combination current constraints from EWPO, DY $d\sigma / dm_{\ell\ell}$ and high invariant mass $A_\text{FB}$, including all terms up to ${\cal{O}}({1/\Lambda^4})$.}
    \label{fig:BmLAndMirrorY}
\end{figure}

\begin{figure}
    \centering
    \includegraphics[width=0.495\textwidth]{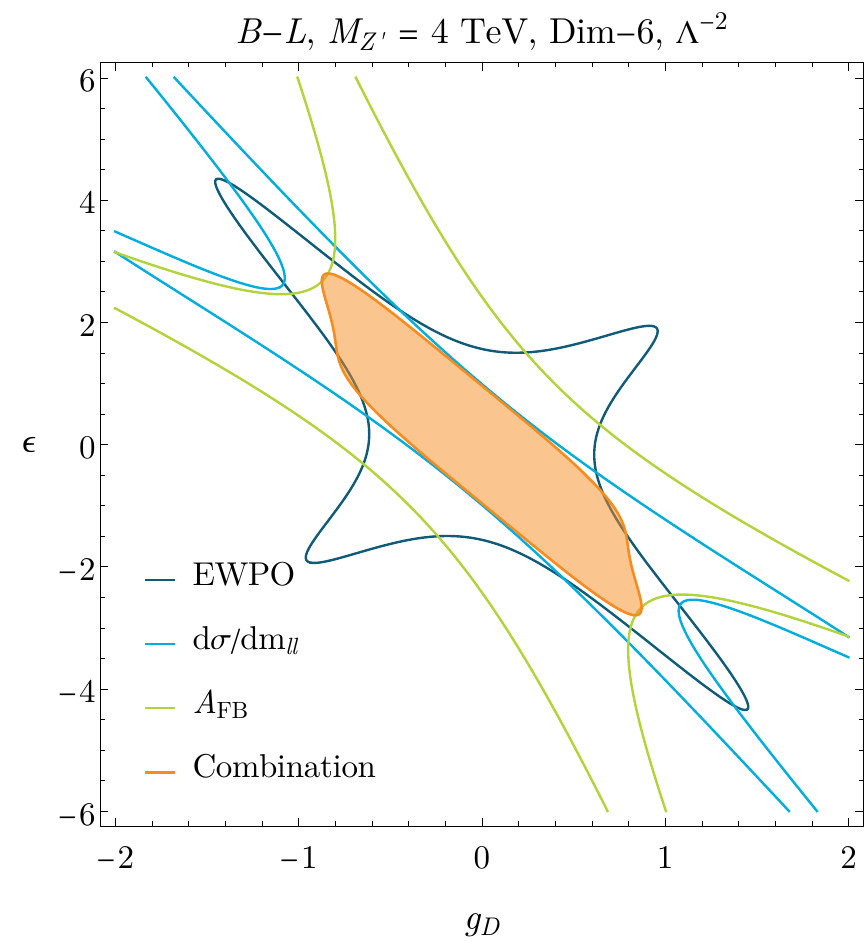}
    \includegraphics[width=0.495\textwidth]{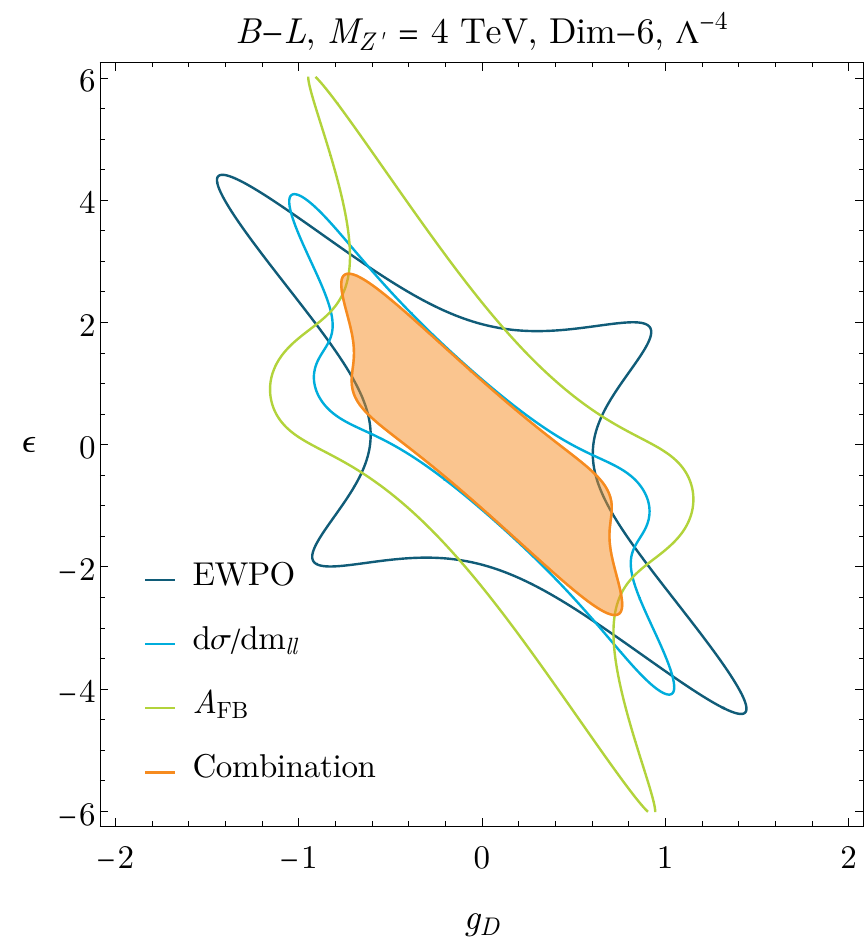}
    \caption{95\% CL constraints in the $(g_D,\epsilon)$ plane for the $B-L$ model with a benchmark mass of $M_{Z'}=4$ TeV when neglecting dimension-8 terms. 
    The left plot shows the constraints including dimension-6 SMEFT contributions up to ${\cal{O}}({1/\Lambda^2})$, while the right also includes (dimension-6)$^2$ ${\cal{O}}({1/\Lambda^4})$ terms.
    We show both separately and in combination current constraints from EWPO, DY $d\sigma / dm_{\ell\ell}$, and high invariant mass $A_\text{FB}$.}
    \label{fig:BmLbreakdown}
\end{figure}

\begin{figure}
    \centering
    \includegraphics[width=0.495\textwidth]{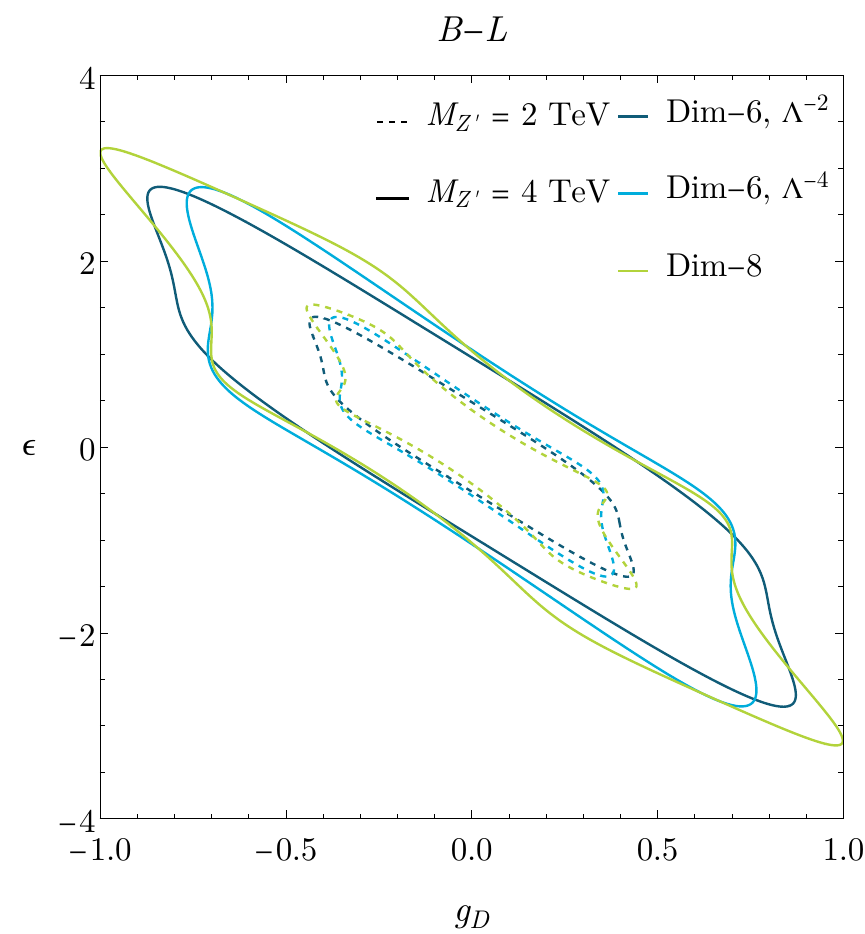}
    \includegraphics[width=0.495\textwidth]{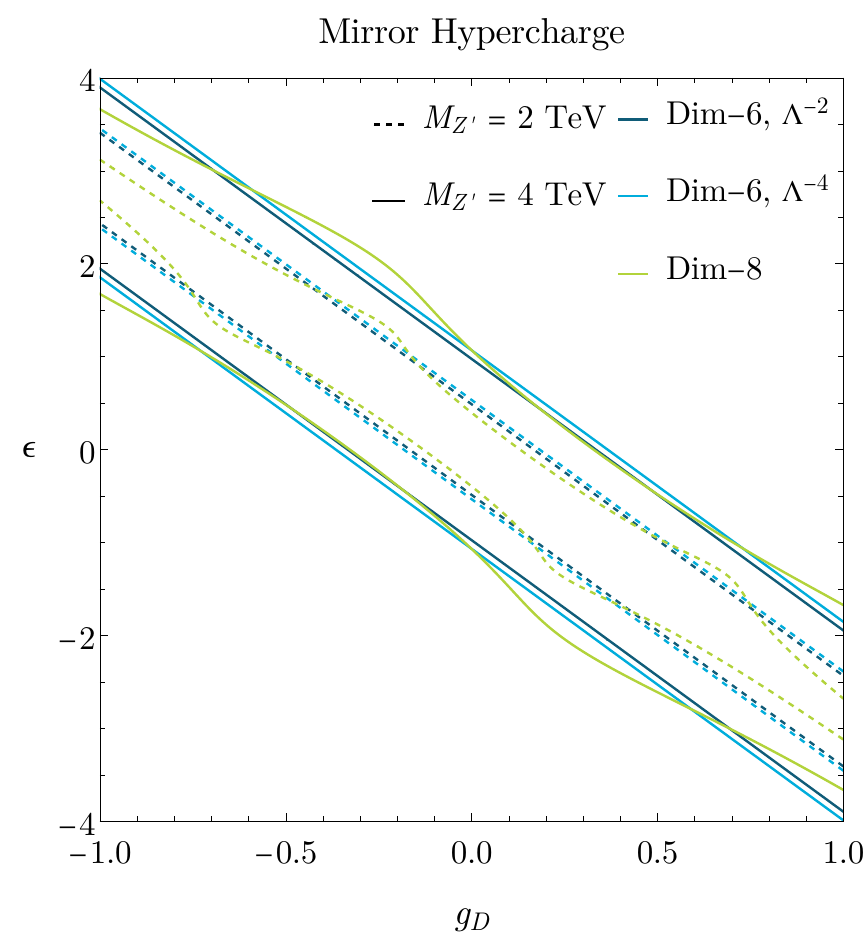}
    \caption{Comparison of the 95\% CL constraints in the $(g_D,\epsilon)$ plane when keeping up to linear dimension-6, (dimension-6)$^2$, and linear dimension-8 contributions to the electroweak precision and Drell-Yan observables. We show results for $B-L$ (left) and mirror hypercharge (right) $Z'$ models for a benchmark mass of $M_{Z^\prime}=2(4)$ TeV as dashed (solid) lines. }
    \label{fig:dim6vdim8}
\end{figure}

Fig.~\ref{fig:BmLAndMirrorY} shows constraints on the $B-L$ and mirror hypercharge models for $M_{Z^\prime}= 4$ TeV,
including all contributions up to ${\cal{O}}({1/\Lambda^4})$.  
For both models, the strongest limits come from measurements of $d\sigma/dm_{\ell\ell}$ and the weakening of the limits for non-zero $\epsilon$ and the flat direction in the mirror hypercharge model when $\mathcal{J}_\mu =-\epsilon j_\mu$  are apparent.
For the $B-L$ case, we show in Fig.~\ref{fig:BmLbreakdown} the same plot when considering only dimension-6 pieces to ${\cal{O}}({1/\Lambda^2})$ and ${\cal{O}}({1/\Lambda^4})$. 
Since the EWPO SMEFT contributions scale as $(m_Z^\text{phys})^2/\Lambda^2$, they are mostly insensitive to the inclusion of ${\cal{O}}({1/\Lambda^4})$ terms. 
On the other hand, the DY constraints exhibit significant changes in shape owing to additional energy enhanced four-fermion operator contributions. 
However, these deviations occur primarily in a region already excluded by EWPOs, and so the combined constraints are accurately captured by using only dimension-6 operators.
This is further illustrated in Fig.~\ref{fig:dim6vdim8} where we show the combined results at two different values of $M_{Z'}$ for the $B-L$ and mirror hypercharge models. 
There are significant shifts when going from $2$~TeV to $4$~TeV, although the linear, ${\cal{O}}({1/\Lambda^2})$, approximation using the dimension-6 operators is sufficient to describe the physics in these models at both mass benchmarks.
As the LHC collects more data and the DY observables become more precise, the inclusion of the ${\cal{O}}({1/\Lambda^4})$ terms will become more important.

\begin{figure}
    \centering
    \includegraphics[width=0.32\textwidth]{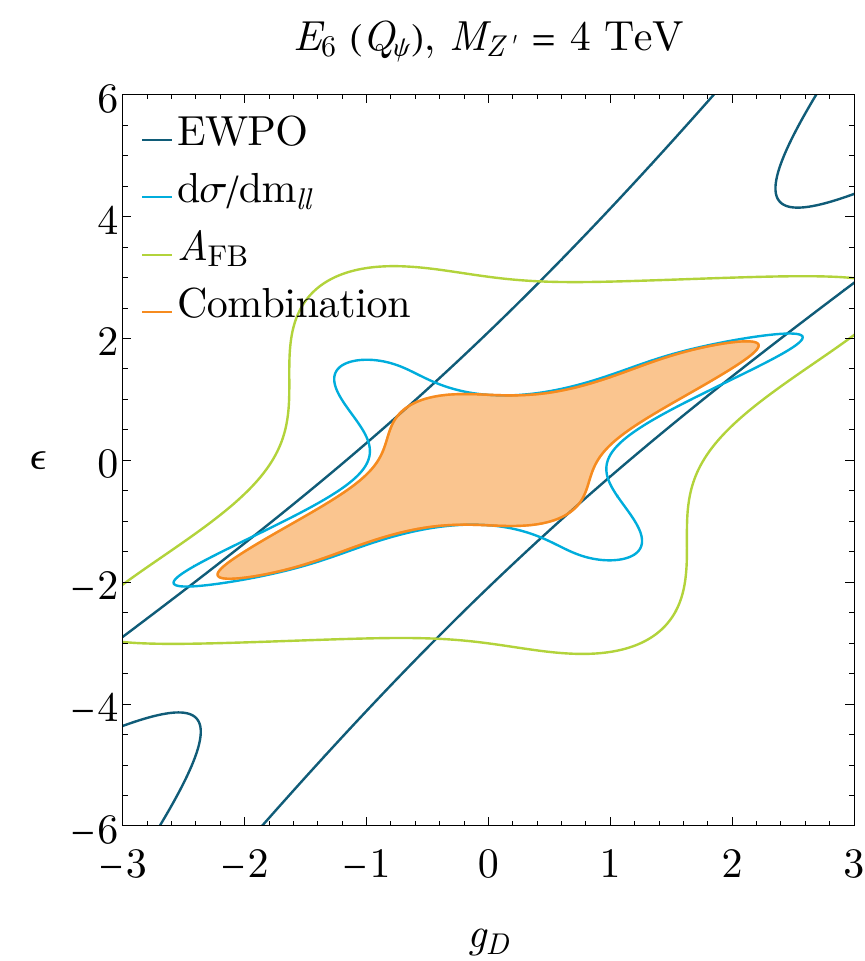}
    \includegraphics[width=0.32\textwidth]{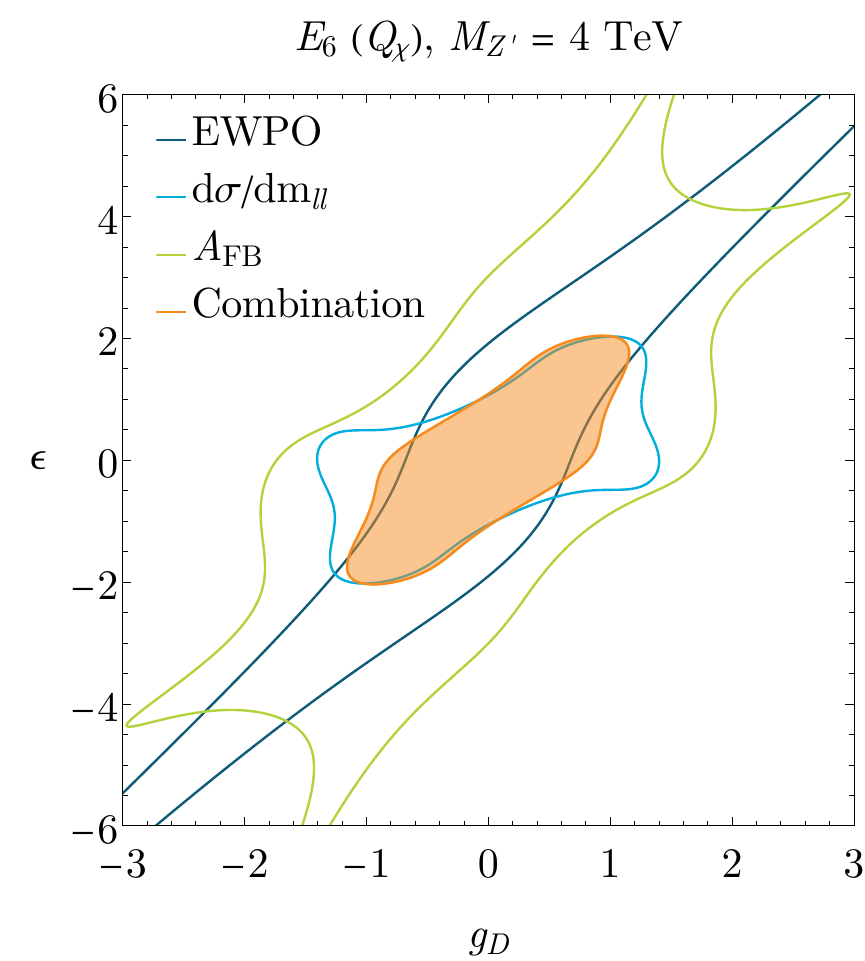}
    \includegraphics[width=0.32\textwidth]{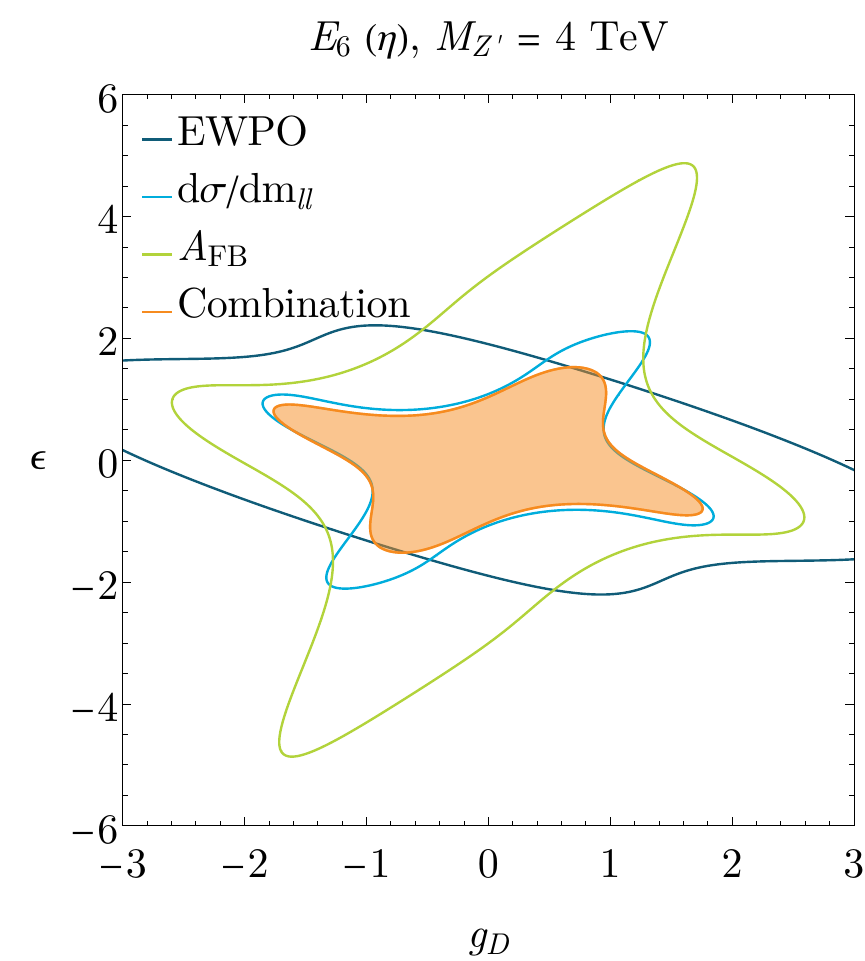}
    \includegraphics[width=0.32\textwidth]{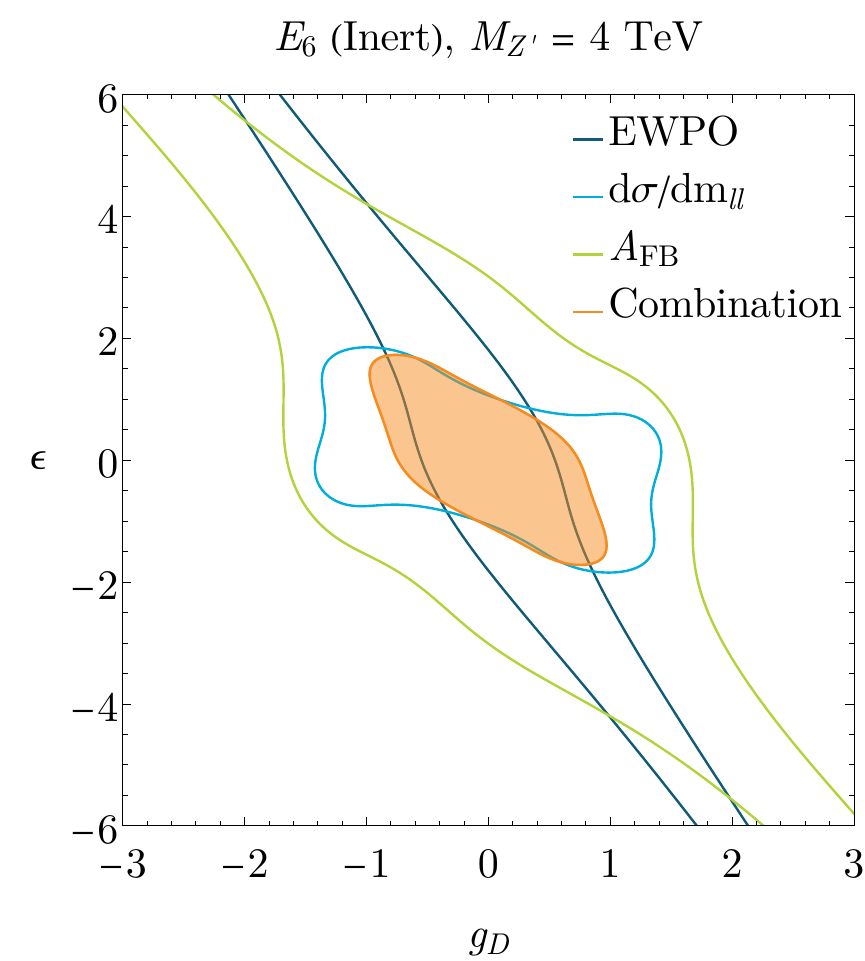}
    \includegraphics[width=0.32\textwidth]{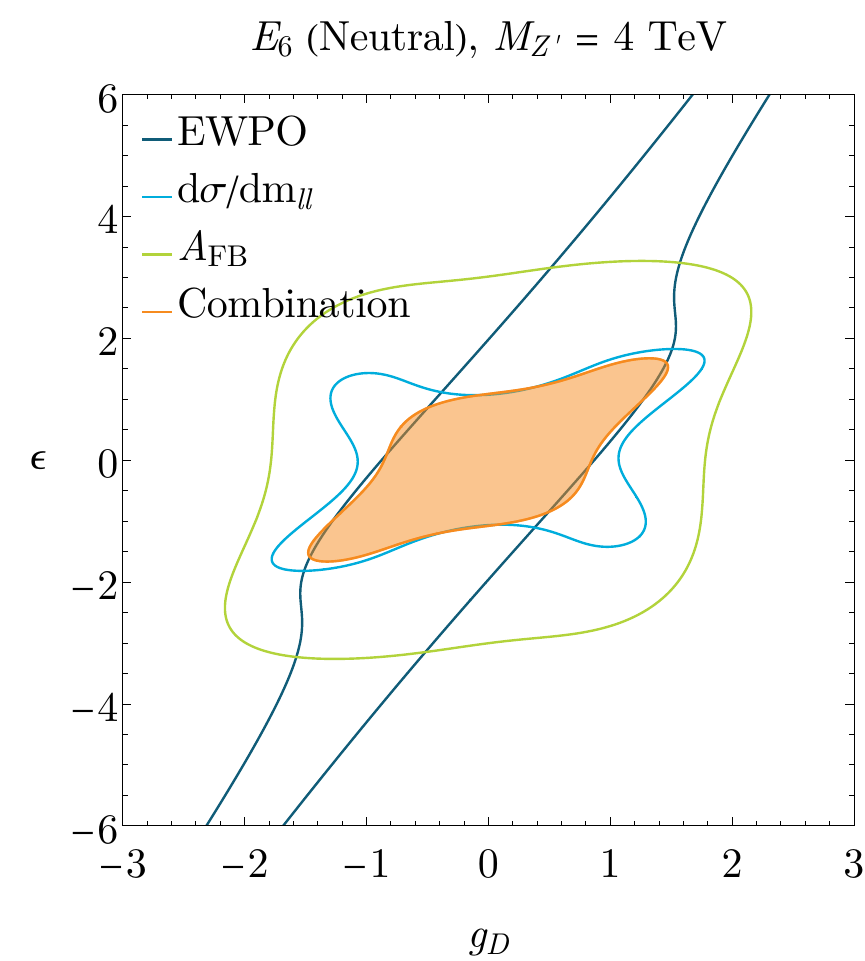}
    \includegraphics[width=0.32\textwidth]{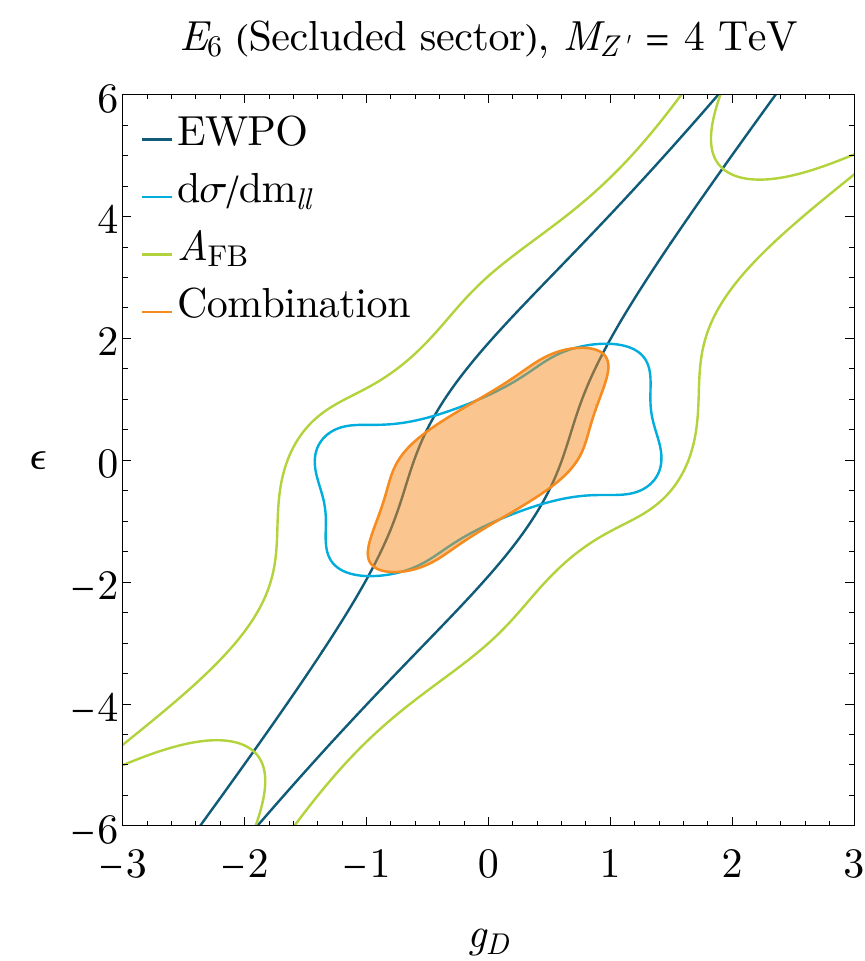}
    \caption{95\% CL constraints in the $(g_D,\epsilon)$ plane for the $E_6$ models we consider with a benchmark mass of $M_{Z'}=4$ TeV. We show both separately and in combination current constraints from EWPO, DY $d\sigma / dm_{\ell\ell}$, and high invariant mass $A_\text{FB}$ including all terms up to ${\cal{O}}({1/\Lambda^4})$.}
    \label{fig:e6constraints}
\end{figure}

\begin{figure}
    \centering
    \includegraphics[width=0.495\textwidth]{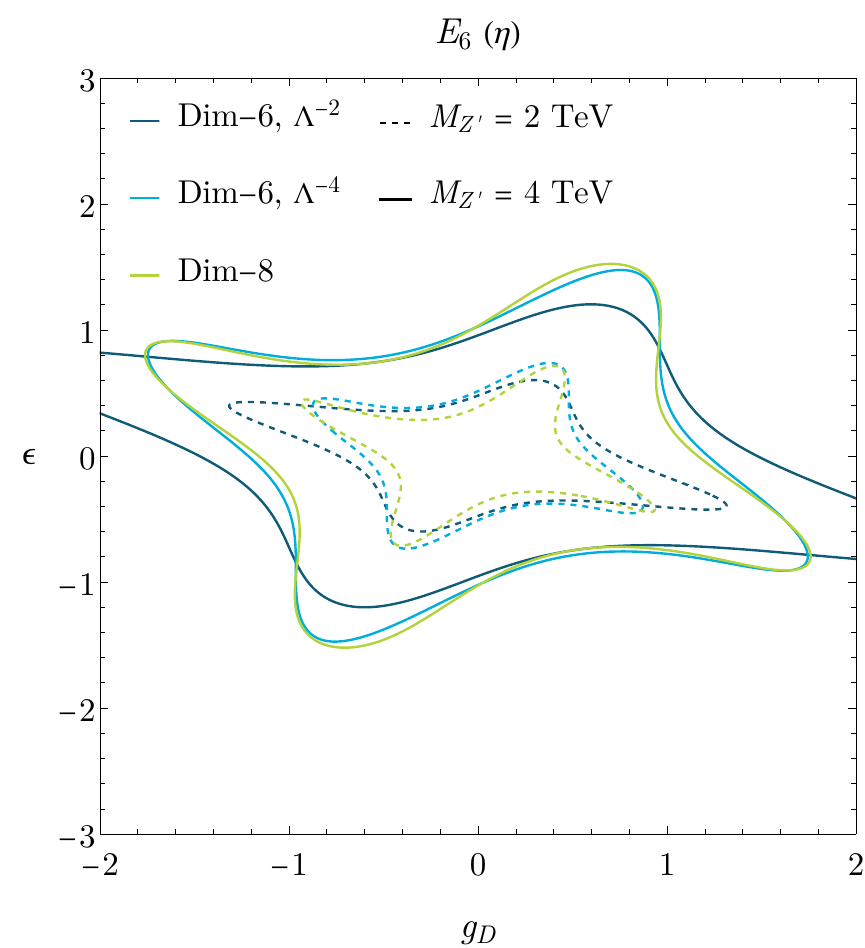}
    \includegraphics[width=0.495\textwidth]{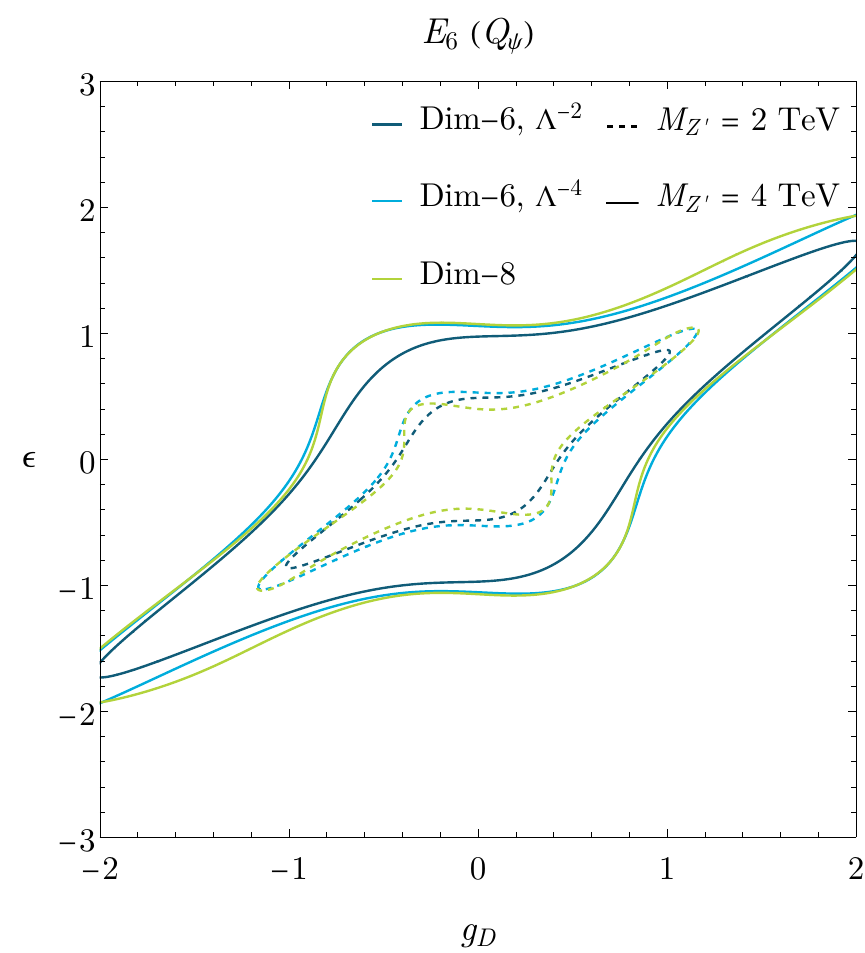}
    \caption{Comparison of the 95\% CL constraints in the $(g_D,\epsilon)$ plane when keeping up to linear dimension-6, (dimension-6)$^2$, and linear dimension-8 contributions to the electroweak precision and Drell-Yan observables. We show results for the $\eta$ (left) and $Q_\psi$ (right) $E_6$ models for a benchmark 
    $Z'$ 
    mass of $M_{Z'}=2(4)$ TeV as dashed (solid) lines. }
    \label{fig:e6D6vD8}
\end{figure}

Fig.~\ref{fig:e6constraints} compares the results in the various $E_6$ scenarios including all terms up to ${\cal{O}}({1/\Lambda^4})$, where the allowed regions span distinct parameter spaces for the different models.   
As before, in all cases, the tightest limits come from $d\sigma/dm_{\ell\ell}$, with $A_\text{FB}$ playing very little role.  
This is to be compared with the results of Ref.~\cite{Boughezal:2023nhe}, which found significant constraints from $A_\text{FB}$ on specific dimension-8 coefficients that are not generated in the $Z^\prime$ models  we consider.  
This suggests that any conclusion about the importance of dimension-8 contributions is model dependent.   
The largest deviations from the dimension-6 only fit for the $E_6$ scenarios are in the cases of the $\eta$ model and the $Q_\psi$ model.
We show the breakdown for these two cases in Fig.~\ref{fig:e6D6vD8}, where the (dimension-6)$^2$ contributions eliminate an approximate flat direction and make a meaningful impact. 
However, dimension-8 contributions are still small corrections, even at $M_{Z'}=2$ TeV. 

\begin{figure}
    \centering
    \includegraphics[width=0.495\textwidth]{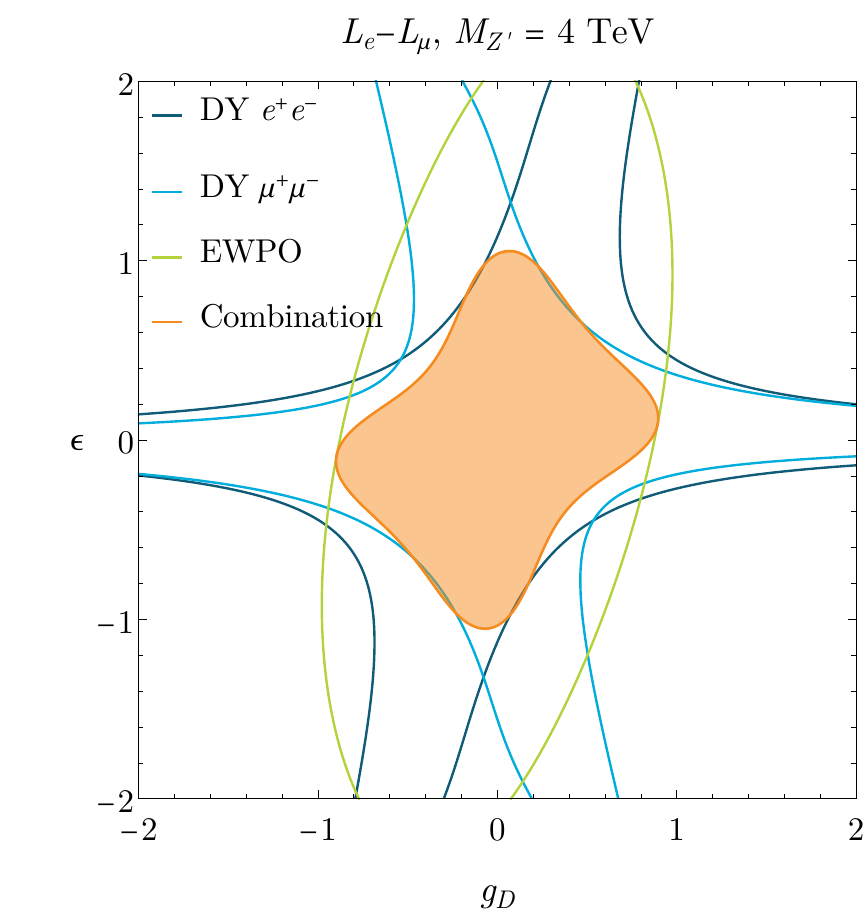}
    \includegraphics[width=0.495\textwidth]{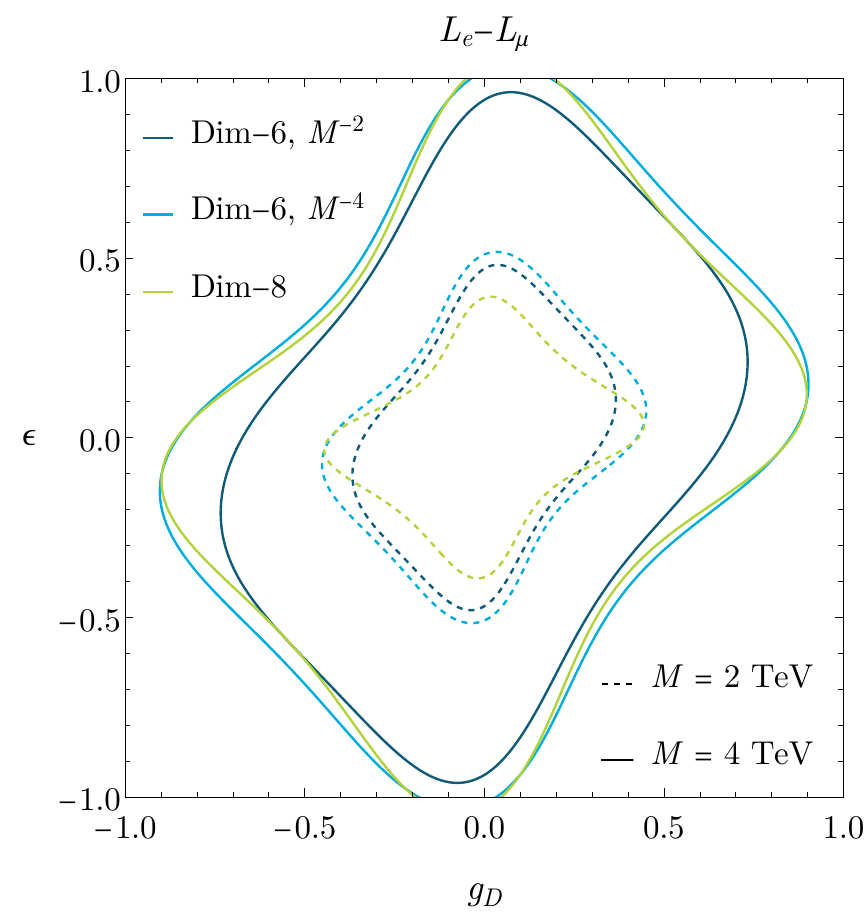}
    \caption{95\% CL constraints in the $(g_D,\epsilon)$ plane for gauged $L_e - L_\mu$. (left) We show current constraints from DY $e^+e^-$ and $\mu^+\mu^-$ production (combining $d\sigma / dm_{\ell\ell}$ and $A_\text{FB}$) and EWPO for a benchmark mass of 4 TeV. (right) For the combination of EWPO's and DY, we show the impact on the constraints when keeping SMEFT contributions up to linear dimension-6, quadratic dimension-6, and linear dimension-8.}
    \label{fig:LEmLMU}
\end{figure}

For an example where we relax flavour universality, we show results for the $L_e-L_\mu$ model in Fig.~\ref{fig:LEmLMU} in the $(g_D,\epsilon)$ plane.
In the left plot, we show for a benchmark mass of 
$M_{Z'}=4$ 
TeV, the breakdown of the limits from EWPOs and DY measurements in the $e^+e^-$ and $\mu^+\mu^-$ channels, respectively, where $A_\text{FB}$ and $d\sigma/dm_{\ell\ell}$ have been combined. 
Since only the leptons are charged under the $U(1)_{L_e-L_\mu}$ in this model, the only contributions to DY have a factor of $\epsilon$, generating a flat direction when $\epsilon=0$ that is bounded by EWPOs from the operator $C_{ll}^{(1)}[1221]$.
The contributions to DY are dominated by contributions from the operators $C_{lq}^{(1)}$, $C_{qe}$, $C_{eu}$, and $C_{ed}$.
For the DY $e^+e^-$ and $\mu^+\mu^-$ channels separately, when $g_D = Q_{e,\mu}g' \epsilon$, the contribution to $C_{qe}$, $C_{eu}$, and $C_{ed}$ cancels, while when $g_D = \frac{1}{2}Q_{e,\mu} g' \epsilon$, the contribution to $C_{lq}^{(1)}$ cancels. 
These cancellations lead to the approximate flat directions seen in the DY $e^+e^-$ and $\mu^+\mu^-$ channels separately, which are removed when combining the two since the charges $Q_{e,\mu}$ of the electrons and muons are different under $U(1)_{L_e-L_\mu}$.
The right plot shows the impact of the (dimension-6)$^2$ and the dimension-8 contributions for 
$M_{Z'}=2$ 
TeV and 
$M_{Z'}=4$ 
TeV, where consistently including up to ${\cal{O}}({1/\Lambda^4})$ leads to a $2\sigma$ constraint on $g_{D}$ up to $\sim$30\% tighter that that from the ${\cal{O}}({1 / \Lambda^2})$ result when $\epsilon=0$, even for 
$M_{Z'}=4$ TeV.
Once again, this shift is driven primarily by the (dimension-6)$^2$ terms, while the dimension-8 terms are a small correction.

\begin{figure}
    \centering
    \includegraphics[width=0.495\textwidth]{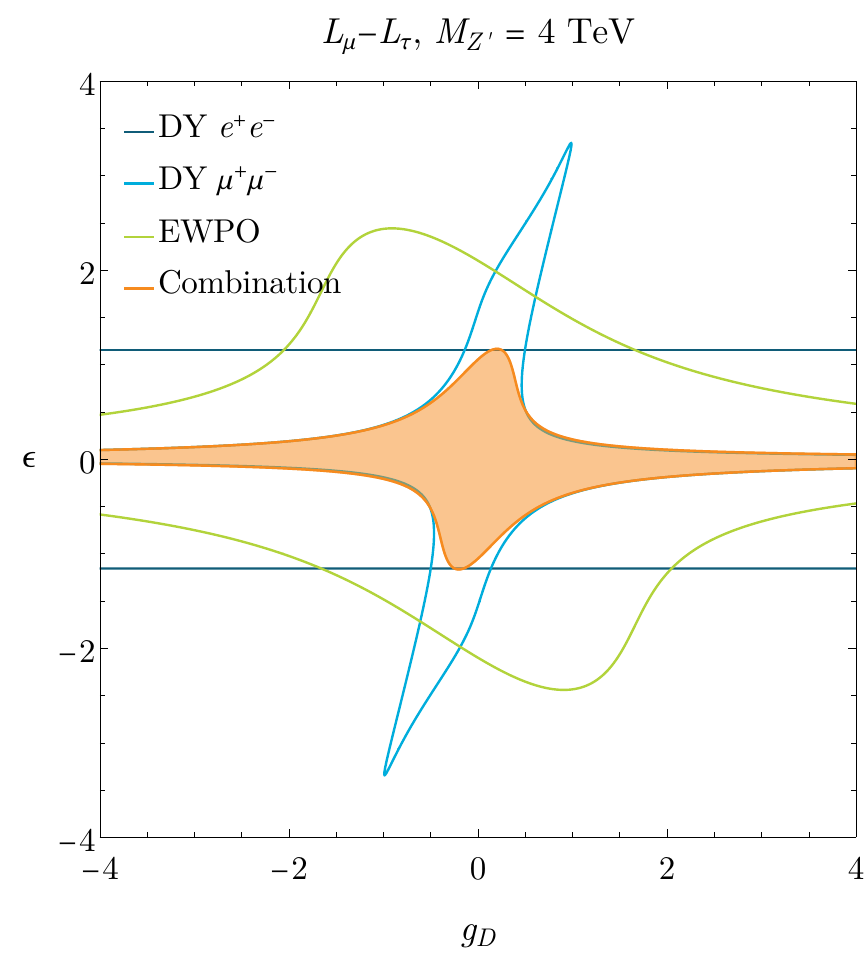}
    \includegraphics[width=0.495\textwidth]{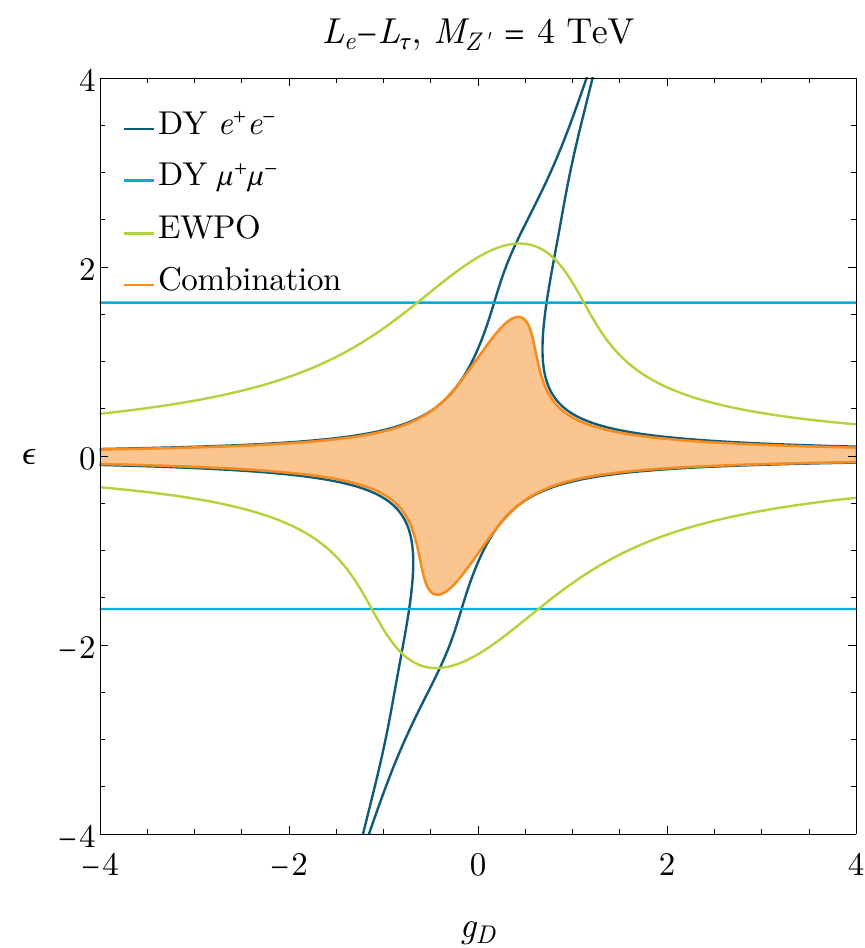}
    \caption{95\% CL constraints in the $(g_D,\epsilon)$ plane for gauged (left) $L_\mu - L_\tau$ and (right) $L_e - L_\tau$.  We show current constraints from DY $e^+e^-$ and $\mu^+\mu^-$ production (combining $d\sigma / dm_{\ell\ell}$ and $A_\text{FB}$) and EWPO for a benchmark mass of 4 TeV.}
    \label{fig:LTAU}
\end{figure}

Results for the other two lepton number difference models, $L_\mu-L_\tau$ and $L_e-L_\tau$, are shown in Fig.~\ref{fig:LTAU}.
In these cases, the operator $C_{ll}^{(1)}[1221]$ is not generated, so $\epsilon=0$ is a true flat direction. 
However, if there is a non-zero $\epsilon$, constraints on $g_{D}$ quickly reemerge, leading to constraints of a similar size as the $L_e-L_\mu$ model. 
Inclusion of DY $\tau^+\tau^-$ data may improve the bound on $\epsilon$, but would not resolve the flat direction when $\epsilon=0$.
Inclusion of additional observables sensitive to lepton flavour such as $\ell^+\ell^- \rightarrow \tau^+ \tau^-$ at a future lepton collider would be necessary to fully remove this flat direction.

\section{Conclusion}
\label{sec:con}
We have considered SMEFT matching to $Z^\prime$ models including all terms of dimension-6 and dimension-8.  
An arbitrary kinetic mixing term is also consistently included and leads to approximate blind directions.  
Restrictions on the model parameters are then derived in the SMEFT framework from EWPOs and from DY $m_{\ell\ell}$ and $A_\text{FB}$ distributions at the LHC.  
In all cases, $A_\text{FB}$ plays little role.  

The impact of the dimension-8 contributions is small, while the $\mathcal{O}(1/\Lambda^4)$ terms from the (dimension-6)$^2$ coefficients have a significant numerical effect, as does the presence of a non-zero kinetic mixing.  
Our limits on a heavy $Z^\prime$ mass range from $2-12$ TeV and demonstrate the model dependence of the SMEFT fits. 
Our results are of the same order of magnitude as SMEFT fits to a heavy $Z^\prime$ with generic couplings to SM particles and illustrate the importance of considering complete UV models for obtaining precise limits.

Of course, the operators we have considered may be probed in other datasets as well. For example, a number of other low-energy probes of four-fermion operators at dimension-6 were studied in~\cite{Falkowski:2017pss}, where they can sometimes be dominant. Low-energy parity violation measurements were shown in~\cite{Boughezal:2021kla} to break some flat directions present when considering Drell-Yan data alone, as well as to disentangle dimension-6 and dimension-8 effects. It would be interesting to see how the $Z'$ model constraints are impacted when including these additional constraints in a global fit.
\acknowledgments
We would like to thank Yingsheng Huang for helpful correspondence on the results in Ref.~\cite{Boughezal:2023nhe} and Tobias Neumann for assistance with MCFM.   
We thank Andrea Thamm for an informative seminar on her results on $Z^\prime$ bosons.  We thank Ben Allenach and Nico Gubernari for pointing out several sign errors in the original version of this paper.
S. D. is supported by  the U.S. Department of Energy under Grant Contract  DE-SC0012704. 
M.F. is supported by the U.S. Department of Energy, Office of Science, Office of Workforce Development for Teachers and Scientists, Office of Science Graduate Student Research (SCGSR) program. The SCGSR program is administered by the Oak Ridge Institute for Science and Education (ORISE) for the DOE. ORISE is managed by ORAU under contract number DE-SC0014664. 
M.S. gratefully acknowledges support from the Alexander von Humboldt Foundation as a Feodor Lynen Fellow. 
Digital data can be found at
\url{https://quark.phy.bnl.gov/Digital_Data_Archive/dawson/zprime_24}. 
\bibliographystyle{utphys}
\bibliography{zprime.bib}
\end{document}